# Providing Secrecy With Structured Codes: Tools and Applications to Two-User Gaussian Channels

Xiang He    Aylin Yener

Wireless Communications and Networking Laboratory

Electrical Engineering Department

The Pennsylvania State University, University Park, PA 16802

*xxh119@psu.edu    yener@ee.psu.edu*

### Abstract

Recent results have shown that structured codes can be used to construct good channel codes, source codes and physical layer network codes for Gaussian channels. For Gaussian channels with secrecy constraints, however, efforts to date rely on random codes. In this work, we advocate that structured codes are useful for providing secrecy, and show how to compute the secrecy rate when structured codes are used. In particular, we solve the problem of bounding equivocation rates with one important class of structured codes, i.e., nested lattice codes. Having established this result, we next demonstrate the use of structured codes for secrecy in two-user Gaussian channels. In particular, with structured codes, we prove that a positive secure degree of freedom is achievable for a large class of fully connected Gaussian channels as long as the channel is not degraded. By way of this, for these channels, we establish that structured codes outperform Gaussian random codes at high SNR. This class of channels include the two-user multiple access wiretap channel, the two-user interference channel with confidential messages and the two-user interference wiretap channel. A notable consequence of this result is that, unlike the case with Gaussian random codes, using structured codes for both transmission and cooperative jamming, it *is possible* to achieve an arbitrary large secrecy rate given enough power.

This work was presented in part at Allerton Conference on Communication, Control, and Computing, September 2008 and submitted in part to IEEE Globecom Conference in March, 2009. This work is supported in part by the National Science Foundation with Grants CCR-0237727, CCF-051483, CNS-0716325, CNS-0721445 and the DARPA ITMANET Program with Grant W911NF-07-1-0028.

## I. INTRODUCTION

The notion of information theoretic secrecy was first proposed by Shannon [1] whereby a message transmitted to a receiver is guaranteed to be kept secret from an eavesdropper, irrespective of the computational power it possesses. In particular, it was shown that it is possible that the eavesdropper gains no information regarding the secret message having intercepted the cryptogram, albeit at the expense of very long keys [1]. Wyner, in [2], established that, more often than not, the eavesdropper (Eve) has a noisy copy of the signal transmitted from the source, and exploiting the channels, building a useful secure communication system per Shannon's notion is possible [2]. Csiszár and Körner [3] extended Wyner's setting to the general discrete memoryless wiretap channel and established its secrecy capacity.

Numerous channel models have since been studied using the information theoretic secrecy framework. The maximum reliable transmission rate with secrecy was identified for some basic models including the Gaussian wiretap channel [4], the MIMO wiretap channel [5], [6] and the MIMO Gaussian broadcast channel with confidential messages [7], [8]. Generally speaking, secrecy capacity regions for multi-transmitter models remain as open problems, though some specific cases are solved, e.g., sum secrecy capacity for a degraded Gaussian multiple access wiretap channel [9], [10]. For other channels, upper bounds, lower bounds and some asymptotic results on the secrecy capacity exist, see for example [11]–[16]. For the achievability part, Shannon's random coding argument proves to be effective in these works.

On the other hand, it is known that the random coding argument may be insufficient to prove capacity theorems for certain channels [17], and that the use of structured codes, e.g., lattice codes, is called for. This body of work suggests that using structured codes has two benefits: First, it is relatively easy to analyze large networks with these codes. For example, in [18], [19], the lattice code allows the relaying scheme to be equivalent to a modulus sum operation, making it easy to trace the signal over a multi-hop relay network. Secondly, the structured nature of these codes makes it possible to align unwanted interference, for example, in the interference channel with more than two users [20]–[23], and the two-way relay channel [18], [19].

A natural question therefore is whether structured codes are *useful* for secret communication as well, that is whether imposing a structure on the codebook carries any *inherent* advantage from the secrecy perspective. In this work, we shall answer this question positively and present a structured coding scheme, for which we will be able to compute the secrecy rate. Furthermore, we will demonstrate that there are channel models where the secrecy rates offered by structured codes outperform that of simple random codes, in particular, for channel models with additive Gaussian noise.

A large body of recent work point to the usefulness of lattice codes for Gaussian channels without secrecy [17]–[19], [21], [22]. However, computing the secrecy rate for lattice codes is especially challenging for Gaussian channels. The challenge arises from the manner in which the channel operates which is natural addition. In contrast, for a modulus channel setting, the analysis and the result follows in a relatively straightforward manner [24]. For example, reference [24] considers the wiretap channel with a cooperative jammer in a modulus channel where the source node uses lattice codes to send the secret message and the cooperative jammer uses lattice codes as a jamming signal to confuse the eavesdropper. The eavesdropper observes the *modulus* sum of the code and the jamming signal, where the sum operation is defined over a finite group. In this setting, the observation of the eavesdropper is independent from the secret message as in the case of the one-time pad in [1], and the secrecy rate computation can be done as shown in [24]. However, as commented in [24], in a Gaussian channel, the eavesdropper receives the sum of the signal from the source and the jamming signal, where the sum operation is over the $N$-dimensional real space rather than over a finite group. The property of independence is therefore lost and the method in [24] no longer applies. While one might attempt to use a brute-force method to compute the secrecy rate for finite $N$, lattice codes in general offer the largest rate when the dimension $N \to \infty$, and this makes designing any such brute-force method impractical.

*The first contribution of this work is to solve the problem of secrecy rate computation when the lattice codes are employed.*

Most lattice codes for power constrained transmission have a similar structure to the one used in [24]. First, a lattice is constructed, which should be a good channel code for the noise/interference. Then, to satisfy the power constraint, the lattice, or its shifted version, is intersected with a bounded set, called the shaping set, to create a set of lattice points with finite average power. Commonly used shaping sets include the sphere or its shell [21], and in the case of *nested* lattice code, the fundamental region of a lattice [25]. The choice of the shaping set, in general, does not result in significant performance difference when there are no secrecy constraints. However, as we show in this work, one shaping set is amenable to secrecy rate analysis. In particular, we prove that using nested lattice codes, the real sum of two lattice points from the same codebook leaks *at most 1 bit of information per channel use* to the eavesdropper regarding the value of one of the lattice points, if the other point is independently generated and uniformly distributed over the codebook. This result allows bounding of the equivocation rate and opens the door for applying nested lattice codes to achieve secrecy in Gaussian channels.

*The second contribution of this work is that we show, for a large class of two-user Gaussian channels with real channel gains, structured codes provide unbounded gains in secrecy rate as compared to randomly generated Gaussian codebooks at high SNR.*

We consider a Gaussian wiretap channel when there is an external transmitter that can help jam the eavesdropper, i.e., a cooperative jammer. This model is significant in that it can be viewed as a special case of many two-user Gaussian channel models with interference and secrecy constraints, including the Gaussian MAC-wiretap channel [12], the Gaussian interference channel with confidential messages [26] and the Gaussian interference channel with an external eavesdropper [27, Section VI], [16]. This channel, which is also called the interference assisted wiretap channel in reference [28], is interesting also because it retains the essence of the problem on the trade-off between secrecy and interference. It is known that introducing interference via a cooperative jamming node into the channel may increase the uncertainty observed by the adversary and hence allow for a higher rate of secret messages [12], [13], [26], [28], [29]. However, to harvest this benefit, the interference should be introduced in an intelligent way such that it is more harmful to the adversary than it is to the intended receiver of the messages. Hence, the key is to achieve a fine balance between secrecy against adversary and harmful interference to the intended receiver. The current state of art is such that the achieved rate is still far from the outer bounds for this model [28], despite recent efforts [30]. For example, the genie outer bound from [28] increases with power $P$ at the speed of $0.5 \log_2(P)$ [30, (69)], while the achievable secrecy rate converges to a constant when $P \to \infty$ [28, Theorem 2]. This means the gap between the achievable rate and the outer bound is unbounded and the trade-off between secrecy and interference is not well-understood. In fact, once the channel model is such that the intended receiver is not harmed by the introduced interference, the achieved secrecy rate immediately comes within $0.5$ bits/channel use of the capacity region, as was shown for the one sided interference channel in [31], the orthogonal MAC wiretap channel in [27] and the two-hop relay channel in [32] when the power of relay is large.

Since the gap between the achievable secrecy rate and its outer bound is most pronounced at high SNR, in this work, we shall use *secure degree of freedom* as the performance measure, which shows the high SNR behavior of the achievable secrecy rate. For a fading interference channel, using interference alignment [33] and fading across the channel states, reference [34] demonstrates that the secure degree of freedom can be made positive. However, in the absence of channel variations, the method in [34] does not apply and as described above, reference [28] shows that the achieved rate using Gaussian codebooks converges to a constant as power increases, which implies the obtained secure degree of freedom with Gaussian signalling is zero. In this work, using structured codes, we prove that a strictly positive degree of freedom is achievable for this case, as long as the channel model is fully connected and not degraded. This means that with the help of only one cooperative jammer, given enough power, an arbitrarily large secrecy rate is achievable for a Gaussian wiretap channel.

In obtaining our results we consider *all* possible channel gain configurations with a *fully* connected channel model. The signaling scheme we find that yields the largest achieved secure degree of freedom calls for a *partial alignment* scheme, in which part of the interference is actually aligned with the intended signal. A final note is that, in this work, for the purpose of completeness, we also utilize integer lattices whenever necessary, as they can achieve a larger secure degree of freedom for a certain set of channel gain configurations of measure zero. This is mainly a consequence of their simple structure, for which the equivocation can be computed precisely, while for the nested lattice code the equivocation can only be bounded.

The rest of the paper is organized as follows. Section II describes the mathematical tool found to bound the secrecy rate when nested lattice codes are used. Section III describes the wiretap channel model with a cooperative jammer. Section IV derives the secure degree of freedom with nested lattice codes. Section V compares it with the secure degree of freedom achieved by integer lattice and points out the cases where integer lattice is useful. Section VI computes the secrecy rate with finite power and demonstrates the advantage of the structured coding scheme over random Gaussian signaling even with moderate SNR. In Section VII, we discuss the effect of imperfect channel state information. Section VIII discusses the channel model with complex channel gains. Section IX concludes the paper. Main proofs are presented in Appendices for clarity.

## II. Results on Nested Lattice Codes

A nested lattice code is defined as an intersection of an $N$-dimensional "fine" lattice $\Lambda$ and the fundamental region of an $N$-dimensional "coarse" lattice $\Lambda_c$, denoted by $\mathcal{V}(\Lambda_c)$. $\Lambda, \Lambda_c \subset \mathbf{R}^N$. The term "nested" comes from the fact that $\Lambda_c \subset \Lambda$. The modulus operation is defined as the quantization error of a point $x$ with respect to the coarse lattice $\Lambda_c$:

$$x \bmod \Lambda_c = x - arg\min_{u \in \Lambda_c} \|x - u\|_2 \tag{1}$$

where $\|x - y\|_2$ is the Euclidean distance between $x$ and $y$ in $\mathbf{R}^N$. It can be verified that $\Lambda \cap \mathcal{V}(\Lambda_c)$ is a finite Abelian group when the addition operation between two elements $x, y \in \Lambda \cap \mathcal{V}(\Lambda_c)$ is defined as

$$x + y \bmod \Lambda_c \tag{2}$$

The signal $X^N$ transmitted over $N$ channel uses from a nested lattice codebook is given by

$$X^N = (u^N + d^N) \bmod \Lambda_c \tag{3}$$

Here $u^N$ is the lattice point chosen from $\Lambda \cap \mathcal{V}(\Lambda_c)$, and $d^N$ is called the dithering vector. Conventionally, $d^N$ is defined as a continuous random vector which is uniformly distributed over $\mathcal{V}(\Lambda_c)$ [25]. We show in Appendix C that for the channel models considered in this work, a fixed dithering vector can be used. Either way, the nature of $d^N$ will not affect the result described below. In the following, we assume $u^N$ is independent from $d^N$. We also assume that $d^N$ is perfectly known by all receiving nodes, and hence, is not used to enhance secrecy.

As will be shown later, our goal in general will be to bound the expression of the form

$$I(u_1^N; X_1^N \pm X_2^N, d_1^N, d_2^N) \tag{4}$$

which will correspond to the rate of information leaked to the eavesdropper. Here $u_i^N, X_i^N, d_i^N$ correspond to the $u^N, X^N, d^N$ mentioned above respectively. That is to say that $u_i^N \in \Lambda \cap \mathcal{V}(\Lambda_c)$; $d_i^N$ is the dithering noise; $X_i^N = (u_i^N + d_i^N) \bmod \Lambda_c$. In addition, $u_i^N, d_i^N, i = 1, 2$ are independent. To bound (4), we start from the following result, which we will term the *representation theorem* from here on:

*Theorem 1:* Let $t_1, t_2, ..., t_K$ be $K$ numbers taken from the fundamental region of a given lattice $\Lambda$. There exists a integer $T$, such that $1 \leq T \leq K^N$, and $\sum_{k=1}^{K} t_k$ is uniquely determined by $\{T, \sum_{k=1}^{K} t_k \bmod \Lambda\}$.

*Proof:* The proof is given in Appendix A. ∎

*Remark 1:* As shown in its proof, Theorem 1 is a purely algebraic result and does not rely on the statistics of $t_1, t_2, ...t_K$. □

When $K = 2$, we have the following corollary:

*Corollary 1:* For $X_i^N, i = 1, 2$ computed according to (3), i.e., $X_i^N = (u_i^N + d_i^N) \bmod \Lambda_c$, there exists an integer $T$, such that $1 \leq T \leq 2^N$, and $X_1^N \pm X_2^N$ is uniquely determined by $\{T, X_1^N \pm X_2^N \bmod \Lambda_c\}$.

*Proof:* Define $-\Lambda_c = \{-x : x \in \Lambda_c\}$. Since $0 \in \Lambda_c$ and the difference of any two lattice points is a lattice point, we have $-\Lambda_c = \Lambda_c$. This means that if $X_2^N \in \mathcal{V}(\Lambda_c)$ , then $-X_2^N \in \mathcal{V}(-\Lambda_c)$. Since $-\Lambda_c = \Lambda_c$, this means that $-X_2^N \in \mathcal{V}(\Lambda_c)$ . Hence the corollary follows from Theorem 1 by letting $t_1^N = X_1^N$ and $t_2^N = \pm X_2^N$. ∎

Using Corollary 1, we have

$$I(u_1^N; X_1^N \pm X_2^N, d_1^N, d_2^N) \tag{5}$$

$$= I(u_1^N; X_1^N \pm X_2^N \bmod \Lambda_c, T, d_1^N, d_2^N) \tag{6}$$

$$\leq I(u_1^N; X_1^N \pm X_2^N \bmod \Lambda_c, d_1^N, d_2^N) + H(T) \tag{7}$$

$$= I(u_1^N; u_1^N \pm u_2^N \bmod \Lambda_c) + H(T) \tag{8}$$

Here the $T$ in (6) is the integer defined in Corollary 1.

Since $\Lambda \cap \mathcal{V}(\Lambda_c)$ is an Abelian group, when $u_2^N$ is independent from $u_1^N$, and $u_2^N$ is uniformly distributed over $\Lambda \cap \mathcal{V}(\Lambda_c)$, we have [24], [35]:

$$I(u_1^N; u_1^N \pm u_2^N \bmod \Lambda_c) = 0 \tag{9}$$

Applying it to (8), we find (5) is upper bounded by

$$H(T) \leq N \tag{10}$$

Equations (6)-(10) imply

$$\frac{1}{N} I\left(u_1^N; X_1^N \pm X_2^N, d_1^N, d_2^N\right) \leq 1 \tag{11}$$

Hence, if the eavesdropper observes $X_1^N \pm X_2^N$ and knows $d_i^N, i = 1, 2$, it can obtain *at most* 1 *bit information per channel use* regarding the value of $u_1^N$, if $u_2^N$ is independently from $u_1^N$ and uniformly distributed over $\Lambda \cap \mathcal{V}(\Lambda_c)$. We will use this result extensively in the sequel.

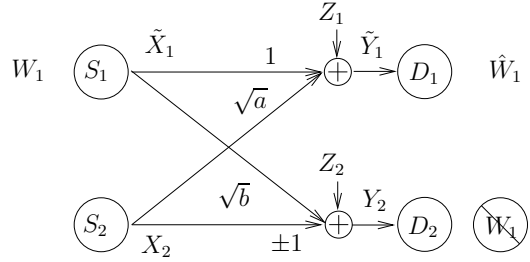

Fig. 1. The Gaussian Wiretap Channel with a Cooperative Jammer

## III. The Gaussian Wiretap Channel with a Cooperative Jammer

### A. System Model and Metric

Consider the Gaussian wiretap channel with a cooperative jammer [28] shown in Figure 1. In this model, node $S_1$ sends a message $W_1$ via $\tilde{X}_1$ to node $D_1$, which must be kept secret from node $D_2$. Node $S_2$, the cooperative jammer, sends signal $X_2$. We assume the channel is fully connected, which means that none of the channel gains equals zero. This assumption is valid for the wireless medium. After normalizing the channel gains of the two intended links to 1, the received signals at the two receiving node $D_1$ and $D_2$ can be expressed as

$$\tilde{Y}_1 = \tilde{X}_1 + \sqrt{a}X_2 + Z_1$$
$$Y_2 = \sqrt{b}\tilde{X}_1 \pm X_2 + Z_2 \tag{12}$$

where $Z_i, i = 1, 2$ is a zero-mean Gaussian random variable with unit variance, and $\sqrt{a}$, $\sqrt{b}$ and $Z_i$ are real numbers.

Let $\hat{W}_1$ be the estimate of $W$, estimated at node $D_1$. For $D_1$ to receive $W_1$ reliably, we require

$$\lim_{n \to \infty} \Pr\left(W_1 \neq \hat{W}_1\right) = 0 \tag{13}$$

In addition, since $W_1$ must be kept secret from $D_2$, we require

$$\lim_{n \to \infty} \frac{1}{n} H\left(W_1\right) = \lim_{n \to \infty} \frac{1}{n} H\left(W_1 | Y_2^n\right) \tag{14}$$

The achieved secrecy rate $R_e$ is defined as:

$$R_e = \lim_{n \to \infty} \frac{1}{n} H\left(W_1\right) \tag{15}$$

such that the conditions (13), (14) are fulfilled simultaneously.

Let $X_1 = \sqrt{b}\tilde{X}_1$ and $Y_1 = \sqrt{b}\tilde{Y}_1$. Then from (12), we have

$$Y_1 = X_1 + \sqrt{ab}X_2 + \sqrt{b}Z_1$$
$$Y_2 = X_1 \pm X_2 + Z_2 \tag{16}$$

In the sequel, we will focus on this scaled model which will be more convenient to explain our results.

There are two constraints on the input distribution to the channel model in (16): First, we assume there is no common randomness shared by the encoders of $S_1$ and $S_2$. This means, the input distribution to the channel is constrained to be

$$p\left(X_1^m, W_1\right) p\left(X_2^m\right) \tag{17}$$

where $m$ is the number of channel uses involved. Intuitively, this implies that if $X_2$ is employed to send interference to confuse the eavesdropper, its effect can not be mitigated by coding $X_1$ via dirty-paper coding [36].

Second, the average power of $X_i$ is constrained to be $\bar{P}_i$. If $X_{i,j}$ is the $j$th component of $X_i$, this means:

$$\lim_{m \to \infty} \frac{1}{m} \sum_{j=1}^{m} E\left[|X_{i,j}|^2\right] \leq \bar{P}_i, i = 1, 2 \tag{18}$$

When computing the secrecy rate $R_e$ for the model in (16), we consider its $n$ symbol extension. For the nested lattice code, $n$ corresponds to the dimension of the lattice $N$. For the integer lattice code, $n = 1$. When we design the coding scheme, we make sure that the signals transmitted by node $S_2$ for every $n$-channel use block are independent from the signals transmitted by it during the other blocks. Hence, if every $n$ channel uses are viewed as a single channel use, the channel is in effect a *memoryless* wiretap channel. This allows us to leverage the following result from reference [3]: Consider a memoryless wiretap channel $\Pr(Y, Z|X)$, where $X$ is the channel input, $Y$ is the observation of the legitimate receiver, $Z$ is the observation of the eavesdropper. Then for a given input distribution $\Pr(X)$, any secrecy rate $R_e$ such that

$$0 \leq R_e \leq [I(X;Y) - I(X;Z)]^+ \tag{19}$$

is achievable. The notation $[x]^+$ equals $x$ if $x \geq 0$ and $0$ otherwise.

Adapting the result to our model, we observe that $X, Y, Z$ corresponds to $X_1^n, Y_1^n, Y_2^n$ respectively. Hence (19) takes the following form:

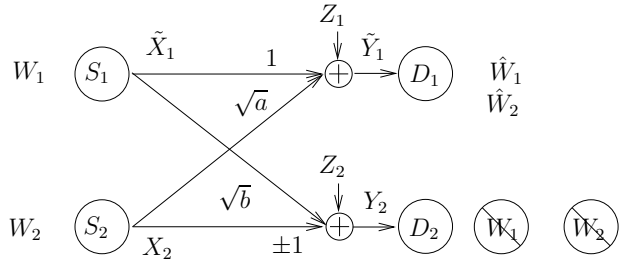

Fig. 2. The Gaussian Wiretap Channel with a Cooperative Jammer as a special case of the Multiple Access Channel with an external eavesdropper, i.e., the MAC-WT [12].

*Theorem 2:* For a given $n$ letter input distribution $\Pr(X_1^n)$, with the assumption that the signals transmitted by node $S_2$ for every $n$-channel use block are independent from the signals transmitted by it during the other blocks, any secrecy rate $R_e$ such that

$$0 \le R_e \le \frac{1}{n}[I(X_1^n; Y_1^n) - I(X_1^n; Y_2^n)]^+ \tag{20}$$

is achievable for the model in (16).

It is important to note that (19) is proved with a random coding argument using joint typicality decoder. However, as we will show later, here the support of the input distribution $\Pr(X_1^n)$ is restricted to the lattice points and hence structure is imposed on the codebook.

*Remark 2:* Note that even though (19) was originally derived for a channel model with discrete alphabets, the same result readily holds for a channel model with continuous alphabets with average power constraint at the transmitter [5, Section IV]. □

In this work, we will mainly be concerned about the high SNR behavior of the secrecy rate. Namely,

*Definition 1:* The secure degree of freedom of the secrecy rate is defined as:

$$\text{s.d.o.f.} = \limsup_{\bar{P}_i \to \infty, i=1,2} \frac{R_e}{\frac{1}{2}\log_2\left(\sum_{i=1}^{2} \bar{P}_i\right)} \tag{21}$$

## B. Relationship with Other Wiretap Channels

The significance of the Gaussian wiretap channel with a cooperative jammer is that it can be considered as a special case of a large class of channel models with secret messages, as explained in the following:

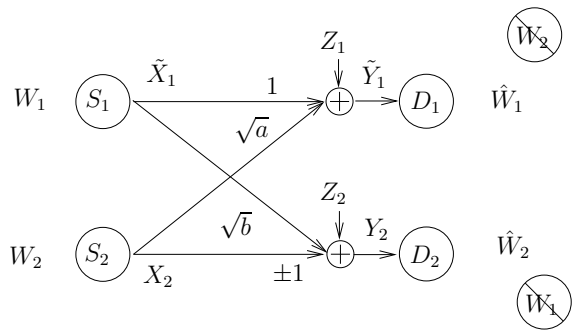

Fig. 3. The Gaussian Wiretap Channel with a Cooperative Jammer as a special case of the two-user Interference Channel with confidential messages, i.e., IFC-CM [26].

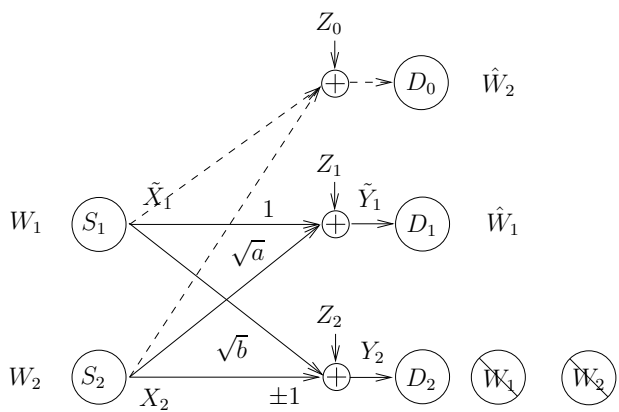

Fig. 4. The Gaussian Wiretap Channel with a Cooperative Jammer as a special case of the Interference Channel with an external eavesdropper, i.e., IFC-WT [27].

1) As shown in Figure 2, if node $S_2$ has a confidential message $W_2$ for $D_1$, which must be kept secret from $D_2$, then the channel is the MAC-wiretap channel [12].

2) If node $S_2$ has a confidential message $W_2$ for $D_2$, and the message must be kept secret from $D_1$, then the channel is the interference channel with confidential messages [26].

3) As shown in Figure 4, we can add another receiving node $D_0$ to Figure 1, to which node $S_2$ wants to sent a confidential message $W_2$. Again $W_2$ must be kept secret from $D_2$. Then the channel becomes the interference channel with an external eavesdropper [27, Section VI] [16].

Hence, it is clear that, any secrecy rate achieved in the Gaussian wiretap channel with a cooperative jammer is an achievable *individual rate* for all the three multi-user channels mentioned above.

An achievable secrecy rate *region* can be obtained by time sharing between these individual rates.

## C. Gaussian Signaling

Reference [28] derived an achievable rate with Gaussian codebooks and power control for the channel in Figure 1. It was shown, in the same reference that, with a Gaussian codebook, the achievable secrecy rate $R_e$ converges to a constant when the power constraints of node $D_1$ and $D_2$ goes to $\infty$. [28, Theorem 2]. This result implies that the achieved secure degree of freedom is 0. It should be noted that in the achievable scheme of [28], node $D_2$ sends a sequence from a codebook randomly generated in an i.i.d. fashion according to a Gaussian distribution. This is the worst noise from the eavesdropper's $(D_2)$ perspective if Gaussian i.i.d. signaling is used in $X_1$, see, for example, [37, Lemma 2]. However, since the channel is fully connected, $X_2$ is also the worst noise for the intended receiver $D_1$. This effect causes the secrecy rate to saturate, leading to zero secure degree of freedom.

A moment's thought reveals that if we allow structure in $X_2$, while the eavesdropper may potentially get more information about $X_1$, the intended receiver $D_1$ may benefit more by exploiting the structure of the interference and neutralizing it more effectively as compared to the eavesdropper. Hence, the overall secrecy rate can be improved. In the following, we will show that this intuition is correct and use structured codes to derive the achievable secure degree of freedom for the Gaussian wiretap channel with a cooperative jammer.

## D. Main result

*Theorem 3:* For the channel model in Figure 1, a positive secure degree of freedom is achievable when the following condition holds:

1) The channel is fully connected. Hence no channel gain is zero.

2) The channel is not degraded [10] or reversely degraded. This means the signal received by the intended receiver can not be expressed as a degraded version, see [10, Section II], of the signal received by the eavesdropper, and vice versa.

The theorem is proved by considering different range of values for $\sqrt{ab}$ in (16). For almost all values of $\sqrt{ab}$, Theorem 3 can be proved using the nested lattice coding scheme, which will be shown in Theorem 4 and Corollary 2. However, there is still a set of channel gains which has measure zero, where the secure degree of freedom achieved by Theorem 4 is zero. For these special cases, we shall use integer lattices to achieve a positive secure degree of freedom. These cases will be discussed in Section V.

## IV. Achievable Secure Degree of Freedom with Nested Lattice Codes

We notice that any $\sqrt{ab}$, $\sqrt{ab} \neq 0$, can be represented in the following form:

$$\sqrt{ab} = p/q + \gamma/q \tag{22}$$

where $p, q$ are positive integers, and $-1 < \gamma < 1, \gamma \neq 0$. In this case, the channel model (16) can be expressed as:

$$qY_1 = qX_1 + (p + \gamma) X_2 + q\sqrt{b}Z_1 \tag{23}$$

$$Y_2 = X_1 \pm X_2 + Z_2 \tag{24}$$

Using this notation, we have the following theorem regarding the achievable secure degree of freedom:

*Theorem 4:* The following secure degree of freedom is achievable using nested lattice codes when $0 < |\gamma| < 0.5$:

$$\left[ \frac{0.25 \log_2 (\alpha) - 1}{\frac{1}{2} \log_2 (\alpha\beta + 1)} \right]^+ \tag{25}$$

where

$$\alpha = \frac{1 - 2\gamma^2 + \sqrt{1 - 4\gamma^2}}{2\gamma^4} \tag{26}$$

and

$$\beta = q^2 + (p + \gamma)^2 \tag{27}$$

*Proof:* The proof is given in Appendix B. ∎

*Remark 3:* In the proof of Theorem 4, we use a layered coding scheme. That is to say that the transmitted signal $X_k^N, k = 1, 2$ is the sum of codewords from $M$ layers as shown in (69), where $X_{k,i}^N$ in (69) is the signal sent by the $S_k$ in the $i$th layer. A nested lattice pair $(\Lambda_i, \Lambda_{c,i}), \Lambda_{c,i} \subset \Lambda_i$, is assigned for each layer. $X_{k,i}^N$ is computed from a lattice point from $\Lambda_i \cap \mathcal{V}(\Lambda_{c,i})$ as shown in (70).

Layered coding scheme with lattice codes was first used in [38] for a $K$-user interference channel ($K \geq 3$) without secrecy constraints. The difference here from [38] is as follows: In [38], a sphere shaped lattice code is used for each layer. Here, for each layer, a nested lattice code is used instead. As a result, the corresponding decoding algorithm and error probability analysis is different. Reference [38] uses the results from [39]. The rate derivation in our work follows reference [25]. □

*Remark 4:* In the proof of Theorem 4, we use the result from [3], i.e., (20), in (97). $X_1^n$ in (20) corresponds to the lattice points from all layers $u_{1,i}^N, i = 1, ..., M$, as shown in (97). □

*Remark 5:* When computing the secrecy rate, we assume the eavesdropper knows (i) the channel noise, as shown by (107), and (ii) the contribution to the received signal from each layer, as shown by (108). For each layer, the information leaked to the eavesdropper takes the form (4). This can be seen from (108) in the proof of Theorem 4. Hence with $M$ layers, the overall information leaked to the eavesdropper is bounded by $M$ bit per channel use, with each layer leaking at most $1$ bit per channel use. □

A consequence of Theorem 4 is as follows:

*Corollary 2:* For $\sqrt{ab}$, such that $2\sqrt{ab}$ is not an integer and $1/\sqrt{ab}$ is not an integer, the secure degree of freedom given by Theorem 4 is positive.

*Proof:* We first verify Corollary 2 holds for interval $1 \leq \sqrt{ab} \leq 2$. The value of (25) is plotted Figure 5 in $1 < \sqrt{ab} < 2$. To prove that (25) is positive in this range, it suffices to choose $(p = 1, q = 1)$, $(p = 2, q = 1)$, $(p = 3, q = 2)$, and let $\sqrt{ab} = p/q + \gamma/q$. A higher secure degree of freedom can be achieved by choosing other values for $p, q$, but for clarity, these curves are not plotted in Figure 5.

Note that for a fixed pair of $p, q$, by changing $\gamma$, (25) takes the shape of a spur. Hence Figure 5 includes three such spurs. The value of (25) converges to $0.5$ when $\gamma$ converges to $0$. However, since $\gamma \neq 0$, the peak of the spur is not included. Hence a positive secure degree of freedom cannot be gauranteed by Theorem 4 only when $\sqrt{ab} = 1, 1.5$ or $2$.

We next argue that Corollary 2 holds for interval $n \leq \sqrt{ab} \leq n+1$ for all integer $n$, $n \geq 1$. This follows from the fact that the denominator of (25) is always positive. Hence the positivity of (25) is only determined by its numerator, which is only a function of $\gamma$. When $n \leq \sqrt{ab} \leq n+1$, we can simply choose the following three pairs of $(p, q)$: $(p = n, q = 1)$, $(p = n + 1, q = 1)$, and $(p = 2n + 1, q = 2)$. The positivity of (25) in this interval $[n, n+1]$ should be the same as the positivity of (25) in interval $[1, 2]$. Since (25) is verified to be positive in interval $[1, 2]$ except for the cases stated in Corollary 2, the same case holds for interval $[n, n + 1]$.

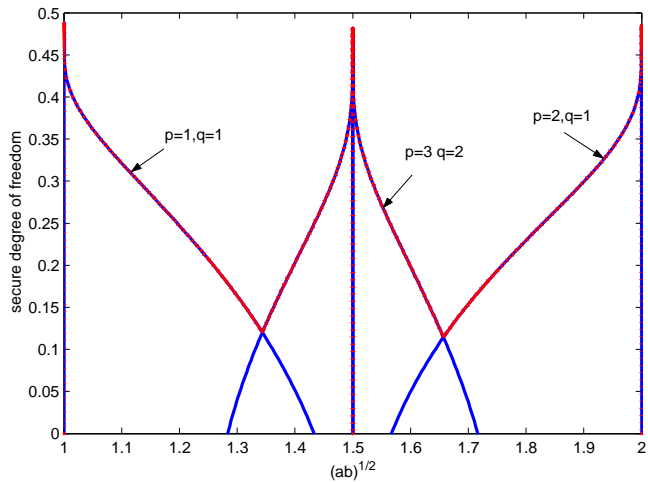

Fig. 5. The value of (25) when $1 < \sqrt{ab} < 2$. The red dashed line is the contour.

We next verify Corollary 2 holds for the interval $1/2 \leq \sqrt{ab} \leq 1$. The value of (25) is plotted in Figure 6. The two spurs in Figure 6 follows from choosing $p = 1, q = 1$, and $p = 1, q = 2$.

Finally we consider the interval $1/(n + 1) \leq \sqrt{ab} \leq 1/n$ for $n \geq 1$. For this interval, we can choose $p = 1, q = n + 1$ and $p = 1, q = n$. Again, since the positivity of (25) is only determined by $\gamma$, Corollary 2 holds for this interval $[1/(n + 1), 1/n]$ since it holds for interval $1/2 \leq \sqrt{ab} \leq 1$.

Hence we have completed the proof of Corollary 2. ∎

*Remark 6:* It should be noted that when $2\sqrt{ab}$ or $1/\sqrt{ab}$ is an integer, and $\sqrt{ab}$ is not equal to one of $\{0.5, 1, 1.5, 2\}$, we can achieve a positive degree of freedom using nested lattice codes also. This entails a different way to align the interference. This alignment scheme is not described in detail because the secure degree of freedom achieved by integer lattice codes is better; see also Remark 9. □

## V. COMPARISON WITH INTEGER LATTICE CODES

Another frequently used class of structured codes is the integer lattice codes [20], [23]. In this section, we compute the secure degree of freedom using integer lattice codes and compare

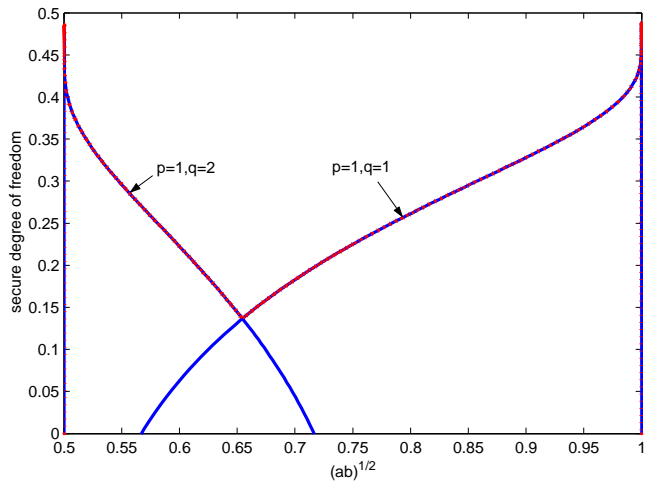

Fig. 6. The value of (25) when $1/2 < \sqrt{ab} < 1$. The red dashed line is the contour.

it with that of the nested lattice codes from Section IV.

### A. Integer Lattice Codes

An integer lattice code with parameter $Q$ is composed of points in the set $[0, Q) \cap \mathbf{Z}$ where $\mathbf{Z}$ is the set of all integers. The point can be scaled and shifted to obtain the actual transmitted signal. Due to its simple structure, unlike nested lattice codes, for this code the rate of information leaked to the eavesdropper can be computed precisely (rather than lower bounded). As will be shown later, the information leaked to the eavesdropper can be bounded by the sum of several terms, each with the following form:

$$f(Q) = I(X_1; X_1 \pm X_2) \tag{28}$$

where $X_i, i = 1, 2$ is uniformly distributed over $[0, Q) \cap \mathbf{Z}$. $f(Q)$ can be bounded by the following lemma:

*Lemma 1:*

$$f(Q) \leq \frac{1}{2} \log_2(2\pi e(\frac{1}{6} - \frac{1}{12Q^2})) < \frac{1}{2} \log_2(\frac{\pi e}{3}) < 0.8 \tag{29}$$

*Proof:* Equation (29) follows directly from [23, Lemma 12]. $H(X_1 + X_2)$ can be bounded as

$$H(X_1 + X_2) \leq \frac{1}{2} \log_2(2\pi e(2P' + \frac{1}{12})) \tag{30}$$

where $P'$ is the variance of $X_k, k = 1, 2$, which is given by

$$P' = \frac{1}{Q} \left[ \sum_{k=0}^{Q-1} k^2 \right] - \left[ \sum_{k=0}^{Q-1} k/Q \right]^2 = \frac{Q^2 - 1}{12} \tag{31}$$

Substituting (31) into (30) we get (29). ∎

As shown by (29), the leakage rate is smaller than the 1-bit bound for the nested lattice code, which suggests that integer lattice codes may be useful for cases where the secrecy rate offered by the nested lattice code is $0$.

*Remark 7:* Compared to the nested lattice code, the drawback of the integer lattice code is its energy inefficiency at finite power. Moreover, typically integer lattices rely on certain number theoretical results. Hence, they only yield good (secure) degree of freedom with very specific channel gain configurations. Luckily, these channel gains happen to be the cases not covered by nested lattice codes in Corollary 2. □

## B. Secure Degree of Freedom

*1) When $\sqrt{ab}$ is algebraic irrational:* When $\sqrt{ab}$ is algebraic irrational, we observe that the achievable secure degree of freedom can be derived using a result from [23]:

*Theorem 5:* A secure degree of freedom of $1/2$ is achievable when $\sqrt{ab}$ is an algebraic irrational number.

*Proof:* The proof is given in Appendix D. ∎

*Remark 8:* When $\sqrt{ab} = 1$ and all channel gains are positive, the channel is degraded and from the outer bound in [28], the secure degree of freedom is 0. Since algebraic irrational numbers are dense on the real line, it follows that the secure degree of freedom is discontinuous at $\sqrt{ab} = 1$. □

Theorem 5 applies only when $\sqrt{ab}$ is algebraic irrational, which is a set of measure $0$ on the real line. Hence other schemes are needed to cover the case where $\sqrt{ab}$ is either rational or transcendental.

*2) When $\sqrt{ab} \geq 2$ or $1/\sqrt{ab} \geq 1/2$:* For this range of $\sqrt{ab}$, we find $Q$-bit expansion scheme similar to the one in [20] is useful in deriving the secure degree of freedom.

*Theorem 6:* Let $Q = \sqrt{ab}$ if $\sqrt{ab} \geq 2$. Otherwise, let $Q = 1/\sqrt{ab}$. Let $\lfloor Q \rfloor$ denotes the largest integer smaller than or equal to $Q$. Let $Q \geq 2$. Then following secure degree of freedom is achievable:

$$\left[ \frac{1}{2} \frac{\log_2 \lfloor Q \rfloor}{\log_2 Q} - \frac{f(\lfloor Q \rfloor)}{2 \log_2 Q} \right]^+ \tag{32}$$

where $f(Q)$ is defined in (28). (32) is lower bounded by

$$\frac{1}{2} \frac{\log_2 \lfloor Q \rfloor}{\log_2 Q} - \frac{\log_2 \left( 2\pi e \left( \frac{1}{6} \right) - \frac{1}{12 \lfloor Q \rfloor^2} \right)}{4 \log_2 (Q)} \tag{33}$$

If $Q = 2$, (32) equals $0.25$.

*Proof:* The proof is given in Appendix E. ∎

In Figure 7, we plot the secure degree of freedom achieved by Theorem 6. The actual performance curve in Figure 7 corresponds to (32). The lower bound curve corresponds to (33). The zigzag shape of the curve is a consequence of the $\lfloor \ \ \rfloor$ operation on $Q$ in (32). We notice as $\sqrt{ab}$ moves away from 1, the lower bound given by (33) becomes tighter, and the secure degree of freedom converges to $0.5$.

In Figure 8, we compare the secure degree of freedom achieved by nested lattice code with those achieved by integer lattice coding scheme described in this section. As shown by Figure 8, neither scheme dominates the other in performance for all channel gains. Nested lattice code offers good secure degree of freedom near the peak of spurs, while integer lattice coding scheme described in this section is at advantage when $2\sqrt{ab}$ are integers and $\sqrt{ab} \geq 2$.

*Remark 9:* A coding scheme similar to the one described in this section can be constructed with a nested lattice code. However, since we can only bound the secrecy rate rather than compute *precisely* the rate of information leaked to the eavesdropper as we did for the integer lattice code, the *provable* secure degree of freedom turns out to be smaller. □

*3) When $\sqrt{ab} = 1$:* When $Y_2 = X_1 + X_2 + Z_2$, the channel is degraded. The secure degree of freedom is known to be $0$ [28].

However, when $Y_2 = X_1 - X_2 + Z_2$, we have the following result:

*Theorem 7:* When $Y_2 = X_1 - X_2 + Z_2$, a secure degree of freedom of $0.0548$ is achievable.

*Proof:* The proof is given in Appendix F. ∎

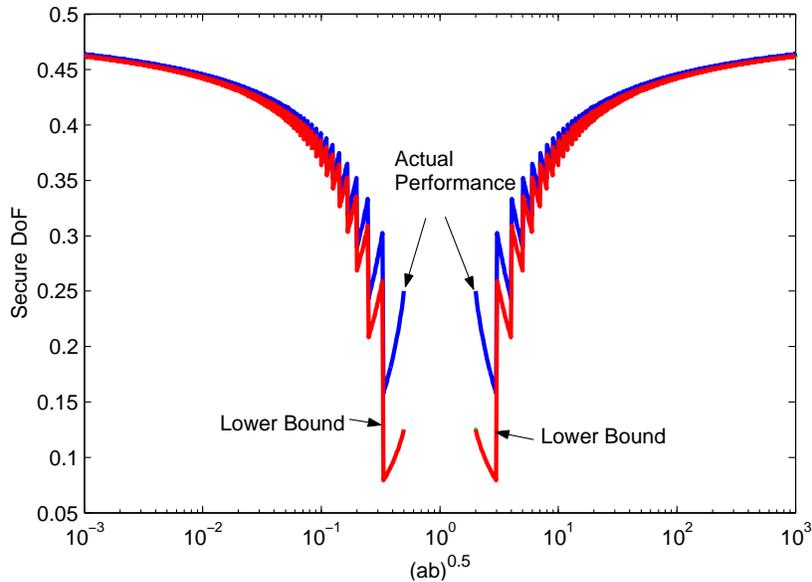

Fig. 7. Secure degree of freedom achieved by integer lattice coding scheme in Section V-B2

*4) When $\sqrt{ab} = 1.5$:* Here we mimic the coding scheme in Section IV when $p = 1, q = 1$. Note that positive secure degree of freedom can not be achieved by the nested lattice code in Section IV in this case. This is mainly a consequence of leaking 1 bit per layer in the layered coding scheme. Here, we replace the nested lattice code with $Q$-bit expansion similar to the one used in Section V-B2 and leverage the fact that integer lattice leaks less than 0.8 bit per layer to show that a positive secure degree of freedom is achievable.

*Theorem 8:* When $\sqrt{ab} = 1.5$, a secure degree of freedom of $1/6$ is achievable.

*Proof:* The proof is given in Appendix G. ∎

*Remark 10:* This scheme can be extended to the case when $\sqrt{ab} = 1 + 1/Q$ with $Q$ being an integer greater than 2. In this case, we let $a_{k,i} = 0$ if $i \bmod 2 = 1$. Otherwise $a_{k,i}$ is taken from $[0, Q-1] \cap \mathbf{Z}$. However, to make room for the carryovers, the least significant bit of the binary representation of $a_{k,i}$ must be zero. Hence, 1 bit is lost per layer due to the carryovers. As a result, the achieved secure degree of freedom turns out to be smaller than those achieved with nested lattice codes. Similar coding schemes can be designed for other values of $\sqrt{ab}$. However, it is difficult to find a uniform description of such codes that achieves a better performance than that of nested lattice code. □

*Remark 11:* At this point, given the description of all these coding schemes using structured

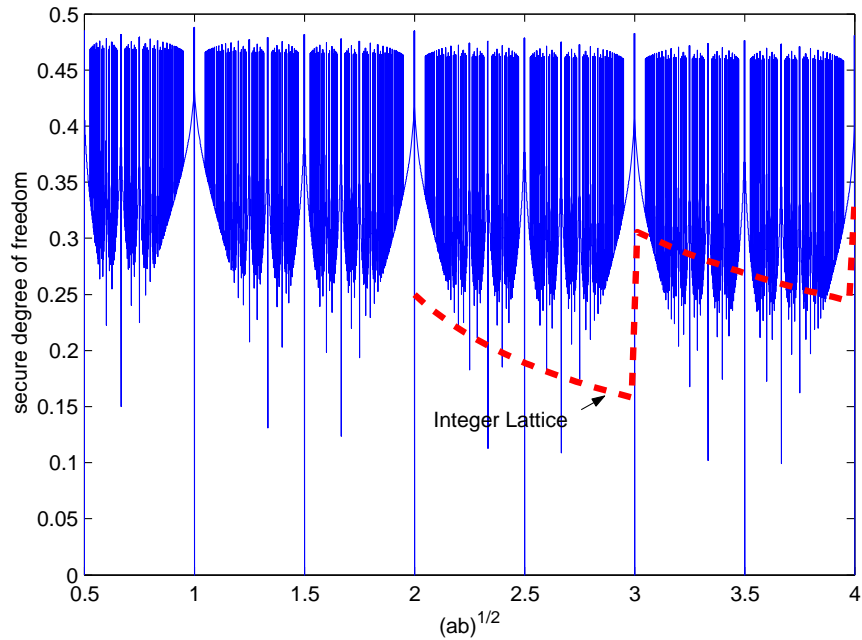

Fig. 8. Comparison of secure degree of freedom achieved by nested lattice code (solid blue) and integer lattice coding scheme in Section V-B2 (dashed red)

codes, the careful reader might find it interesting to ask whether a Gaussian signaling scheme can be devised by mimicking these schemes. We notice this would be difficult for the schemes in Section V-B1 or Section IV, since Section V-B1 relies on [23] which uses results on Diophantine approximation, and Section IV uses partial interference alignment which is difficult for codes without a structure. We are then left with the schemes in Section V-B2 and Section V-B3.

In Sections V-B2 and V-B3, we use multi-level coding schemes. When computing the secrecy rate, it is assumed the eavesdropper has perfect knowledge of the channel noise $Z_2$ and the part of received signals *from each layer*. Note that if a random codebook is used in this model for each layer, revealing $Z_2$ to the eavesdropper would essentially reveal to the eavesdropper the codeword used at each layer. This can be seen by observing that the capacity region of a 2-user Gaussian MAC channel will expand to the whole two dimensional plane if the variance of the channel noise goes to $0$. On the other hand, if $Z_2$ is not revealed to the eavesdropper, then it would be difficult to compute the equivocation rate if a multi-level coding scheme is used. Suppose, to solve this problem, we allocate a small amount of power at each layer to send noise. Doing so would then cause problems at the intended receiver, since the noise can not be decoded and subtracted as the decoder proceeds from higher layers to lower layers. Therefore, we conclude, even if a Gaussian signaling scheme could be eventually found, it would not be a straightforward extensions from any of the coding schemes described in this paper. □

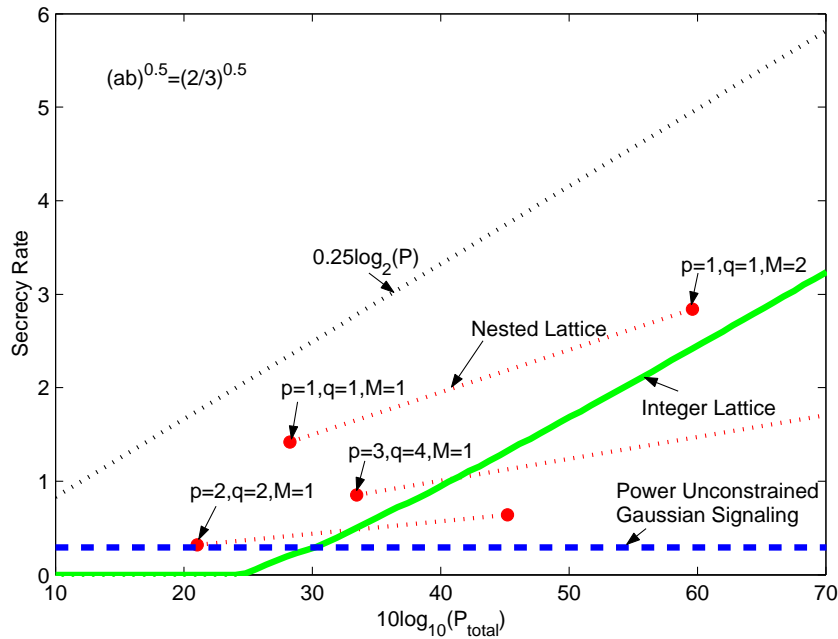

Fig. 9. Secrecy Rate with Finite Power, $a = b = \sqrt{\frac{2}{3}}$, $P_{total}$ is the power of $X_1$ or $X_2$ in (16)

## VI. FINITE POWER SCENARIO

In practice, it is also interesting to look at how fast the asymptotic behavior, i.e., the secure degree of freedom, kicks in as the power increases. In Figure 9, we plot the secrecy rate against power when $a = b = \sqrt{\frac{2}{3}}$. The power $P_{total}$ is the variance of $X_1$ or $X_2$ in (16).

Since $\sqrt{ab} = \sqrt{\frac{2}{3}}$ is algebraic irrational, the integer lattice coding scheme from Section V-B1 provides the largest secure degree of freedom, which is $0.5$. We use (134) to compute its secrecy rate.

From (31) and (125), the average power of this coding scheme is given by

$$P_{total} = P^{1/2+2\varepsilon}\frac{Q^2 - 1}{12} \tag{34}$$

The first term in (34) is due to the scaling factor $P^{1/4+\varepsilon}$ in (125). We then choose different values for $\varepsilon$ and plot the largest achievable power rate pair region in terms of $\{10 \log_{10} P_{total}, R_e\}$ in Figure 9.

For comparison, we plotted the largest secrecy rate offered by the Gaussian random codebook when the power constraints go to $\infty$. Recall that when the power constraints go to $\infty$, the secrecy rate is not affected by whether the channel is described by (12) or (16), which only affects the power constraints. To leverage the result from [28], we use the channel model description in (12). Since $a = b = \sqrt{2/3} < 1$, (12) corresponds to the second case in [28, (11)][1]. The power unconstrained secrecy rate offered by the Gaussian random codebook is given by

$$\lim_{\bar{P}_k \to \infty, k=1,2} R_e = \frac{1}{2} \log_2 \frac{1}{ab} \tag{35}$$

As shown by Figure 9, the secrecy rate offered by integer lattice is greater than the power unconstrained Gaussian signaling scheme when $10 \log_{10} P_{total} > 30$.

We also plot, in Figure 9, the secrecy rate offered by the nested lattice code, by choosing different $p$, $q$ and the number of layers $M$. The secrecy rate is computed according to (111). $P_{total}$ is given by (96). As expected, different choices of $p$, $q$ results in different slope of the curve, hence different secure degree of freedom. When $10 \log_{10} P_{total} < 60$, the nested lattice code can achieve a larger secrecy rate than integer lattice code due to its power efficiency. However, if $P_{total}$ keeps increasing, the secrecy rate offered by the integer lattice is the largest, due to its high secure degree of freedom.

To summarize, these results demonstrate that the coding scheme presented in this paper can outperform random Gaussian signaling at *practical* SNR values as well.

## VII. Channel Gain Mismatch

The coding schemes in previous sections require aligning lattice points at the eavesdropper, which relies on accurate channel state information. It is conceivable that in practice such accurate channel state information may be difficult to obtain and it is reasonable to ask if the nested lattice coding scheme is still able to provide secrecy rate with imperfect channel state information. In this section, we explain how to compute the secrecy rate in that case.

The channel model is shown in Figure 10. This is the same channel model we have been using except now, the channel gain between $S_1$ and $D_2$ has an estimation error $\alpha_i$ for the $i$th channel use. For simplicity, we assume the remaining channels are estimated perfectly. The channel estimation error, $\alpha_i$, is independent from the signal transmitted by node $S_1$, and $\alpha_i \in [-\alpha_{\max}, \alpha_{\max}]$. $\alpha_{\max} > 0$. We also assume the eavesdropper, node $D_2$, has perfect knowledge of $\alpha_i$, but the other nodes only know $\alpha_{\max}$. In this setting, perfect channel state information at the eavesdropper is obviously a pessimistic assumption, but it is insightful in that it will reveal a worst case performance.

---

[1] [28] only considers the case $Y_2 = X_1 + X_2 + Z_2$. However, since the Gaussian distribution is symmetric around zero, the case $Y_2 = X_1 - X_2 + Z_2$ has the same secret rate with Gaussian input distributions.

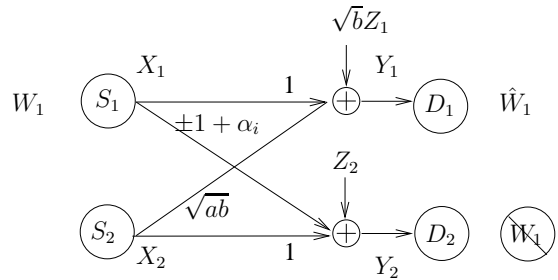

Fig. 10. Wiretap Channel with a Cooperative Jammer $S_2$, where the channel gain between $S_1$ and $D_1$ has an estimation error $\alpha_i \in [-\alpha_{\max}, \alpha_{\max}]$ for the $i$th channel use. $\alpha_i$ is known by $D_2$ but not known by $S_1$ or $S_2$.

With these assumptions, we have the following theorem:

*Theorem 9:* Let $P_1$ be the total power consumption of $S_1$. If for a constant $c > 0$,

$$\alpha_{\max}^2 \leq \frac{c}{P_1} \tag{36}$$

Then the *same* secure degree of freedom given by Theorem 4 is achievable.

*Proof:* The proof is given in Appendix H. ∎

*Remark 12:* Imperfect channel state information for other links, for example, on the link between legitimate transmission pair $S_1$ and $D_1$ can be easily catered for, by using the existing results for the model without secrecy constraints [40, Section IV]. □

## VIII. Complex Channel Gains

The role of complex channel gains in interference alignment has recently been considered [41]. For the purpose of completeness, in this section, we consider the channel model with complex channel gains. We demonstrate that, in this case, by taking advantage of the phase difference existing in the complex channel model, it is easier to achieve positive secure degree of freedom using simple random coding arguments.

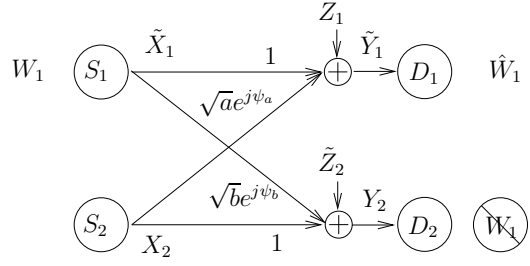

Fig. 11.   The Gaussian Wiretap Channel with a Cooperative Jammer with complex channel gains

The complex Gaussian wiretap channel with a cooperative jammer is shown in Figure 11. Since the channel is fully connected, after normalization, the signals received by the two nodes $D_1$ and $D_2$ can be expressed as

$$\tilde{Y}_1 = \tilde{X}_1 + \sqrt{a}e^{j\psi_a}X_2 + \tilde{Z}_1$$
$$Y_2 = \sqrt{b}e^{j\psi_b}\tilde{X}_1 + X_2 + \tilde{Z}_2 \tag{37}$$

where $j = \sqrt{-1}$. $\tilde{Z}_i, i = 1, 2$ is a rotation invariant complex Gaussian random variable with unit variance. This means $E[|\tilde{Z}_i|^2] = 1$. Let $X_1 = \sqrt{b}e^{j\psi_b}\tilde{X}_1$ and $Y_1 = \sqrt{b}e^{j\psi_b}\tilde{Y}_1$. Let $\psi = \psi_a + \psi_b$. $Z_1 = e^{j\psi_b}\tilde{Z}_1$. $Z_2 = \tilde{Z}_2$. Then from (37) we have

$$Y_1 = X_1 + \sqrt{ab}e^{j\psi}X_2 + \sqrt{b}Z_1$$
$$Y_2 = X_1 + X_2 + Z_2 \tag{38}$$

The input distribution to this channel is constrained to be of the form given in (17). The power constraints of this channel are given by (18). The secrecy rate and secure degree of freedom is defined as in Section III-A. Given these, we have the following theorem:

*Theorem 10:* A secure degree of freedom of 1 is achievable if $\psi \neq 0$ or $\pi \mod 2\pi$

*Proof:* Let $\text{Im}X_i = 0, i = 1, 2$. Let $\cot x = \cos x / \sin x$. Then since $\text{Im}Y_2 = \text{Im}Z_2$, $\text{Im}Y_2$ does not provide any information about $W_1$ to the eavesdropper. Hence we can assume the eavesdropper receives $\text{Re}Y_2$ only. Node $D_1$ computes $g(Y_1) = \text{Re}Y_1 - \cot\psi\text{Im}Y_1$. Then the channel can be expressed as

$$g(Y_1) = \text{Re}X_1 + \sqrt{b}\left(\text{Re}Z_1 - \cot\psi\text{Im}Z_1\right) \tag{39}$$

$$\text{Re}Y_2 = \text{Re}X_1 + \text{Re}X_2 + \text{Re}Z_2 \tag{40}$$

By transmitting i.i.d. Gaussian noise via $\mathrm{Re}X_2$, the channel is equivalent to a Gaussian wiretap channel. It is thus known that the following secrecy rate is achievable [4]:

$$\left[ C\left( \frac{P_1}{(b\csc^2\psi)/2} \right) - C\left( \frac{P_1}{P_2 + 1/2} \right) \right]^+ \tag{41}$$

where $(b\csc^2\psi)/2$ is the variance of the noise term $\sqrt{b}\left(\mathrm{Re}Z_1 - \cot\psi\mathrm{Im}Z_1\right)$. $C(x) = \frac{1}{2}\log_2(1 + x)$. $P_i$, $i = 1,2$, is the average power consumed by node $S_i$. Let $P_i = \bar{P}_i, i = 1,2$ and apply the definition in (21), we observe that a secure degree of freedom of 1 is achievable for this channel. ∎

*Remark 13:* It is understood that if $\log_2\left(\sum_{i=1}^{2}\bar{P}_i\right)$ is used instead of $\frac{1}{2}\log_2\left(\sum_{i=1}^{2}\bar{P}_i\right)$ in the definition of secure degree of freedom (21), the achieved degree of freedom should be changed accordingly from 1 to 0.5. □

*Remark 14:* Since the set of complex numbers is isomorphic to the set of scaled $2 \times 2$ orthogonal matrix, the result here can be interpreted as Gaussian codes being able to achieve positive secure degree of freedom for a wiretap channel with a cooperative jammer where each node has 2 antennas, the channel has real channel gains, and the channel matrix is a scaled orthogonal matrix. □

## IX. Conclusion

Structured codes were shown recently as a useful technique to prove information theoretic results. In this work, we showed that structured codes are also useful to prove secrecy results. We first provided an analytical tool that is useful in computing secrecy rate for nested lattice codes. Then, using integer lattice codes and nested lattice codes, we prove that a positive secure degree of freedom is achievable for the Gaussian wiretap channel with a cooperative jammer and real channel gains as long as the channel is not degraded and the channel is fully connected. Since the channel is a special case of the MAC wiretap channel, the two-user interference channel with confidential messages, and the two-user interference channel with an external eavesdropper, the result means the secure degree of freedom for all these channels are positive as well as long as the channel is not degraded and is fully connected. As a consequence of this high SNR result, we are able to claim that, Gaussian signaling is *not* optimal for a large class of two-user Gaussian channels.

The result here provides further evidence that structured codes are useful. A list of examples that structured codes outperform simple random coding arguments in non-secrecy problems can be found in [17]. Employing structured codes in secrecy problems was first proposed by the authors [42]. Up to date structured codes are found to be useful for relay channels due to the possibility of compute-and-forward [17], [42], or for interference channels with more than two users due to the possibility of interference alignment [23], [38], [43]. The result here provides the example that structured codes are useful for two-user Gaussian channels as well.

An added practical value of our result is that it implies that the cooperation of just one node is sufficient to achieve an arbitrarily large secrecy rate given enough power. Since large scale cooperation involving multiple nodes is not essential, this fact enhances the robustness of the network in an adverse environment.

## APPENDIX A
## PROOF OF THEOREM 1

Let $\mathcal{V}$ be the fundamental region of the $N$-dimensional lattice $\Lambda$. For any set $A$, define $\alpha A$ as $\alpha A = \{\alpha x : x \in A\}$. Then we have:

$$\{\sum_{k=1}^{K} t_k : t_k \in \mathcal{V}, k = 1...K\} = K\mathcal{V} \tag{42}$$

By definition of the modulus $\Lambda$ operation, we have

$$\sum_{k=1}^{K} t_k \bmod \Lambda = \sum_{k=1}^{K} t_k + x, \quad x \in \Lambda \tag{43}$$

Therefore, the theorem is equivalent to finding the number of possible $x$ in equation (43) for a given $\sum_{k=1}^{K} t_k \bmod \Lambda$ such that $t_k \in \mathcal{V}, k = 1...K$. This, by (42), implies:

$$\sum_{k=1}^{K} t_k \bmod \Lambda - x \in K\mathcal{V}, x \in \Lambda \tag{44}$$

To do that, we need to know a little more about the structure of lattice $\Lambda$. Each point in a lattice, by definition, can be represented in the following form [44]: $x = \sum_{i=1}^{N} a_i v_i, \quad v_i \in R^N, a_i \in Z$. $\{a_i\}$ is said to be the coordinates of the lattice point $x$ under the basis $\{v_i\}$.

Based on this representation, we can define the following relationship: Consider two points $x, y \in \Lambda$, with coordinates $\{a_i\}$ and $\{b_i\}$ respectively. Then we say $x \sim y$ if $a_i = b_i \bmod K$, $i = 1...N$. It is easy to see the relationship $\sim$ is an equivalence relationship. Therefore, it defines a partition over $\Lambda$. A partition defined in this way has the following property:

1) Depending on the values of $a_i - b_i \mod K$, there are $K^N$ sets in this partition.

2) The sub-lattice $K\Lambda$ is one set in the partition. The remaining $K^N - 1$ sets are its cosets. Let $C_i$ denote any one of these cosets or $K\Lambda$. Then $C_i$ can expressed as $C_i = K\Lambda + y_i$, $y_i \in \Lambda$. It is easy to verify that $\{x + K\mathcal{V}, x \in C_i\}$ is a partition of $K\mathcal{R}^N + y_i$, which equals $\mathcal{R}^N$.

We now return to solve (43). Since $C_i, i = 1...K^N$ is a partition of $\Lambda$, (43) can be solved by considering the following $K^N$ equations:

$$\sum_{k=1}^{K} t_k \mod \Lambda - x \in K\mathcal{V}, \quad x \in C_i \tag{45}$$

From (42), this means $\sum_{k=1}^{K} t_k \mod \Lambda \in x + K\mathcal{V}$ for some $x \in C_i$. Since $\{x + K\mathcal{V}, x \in C_i\}$ is a partition of $\mathcal{R}^N$, there is exactly one $x \in C_i$ that meets this requirement. This implies for a given $\sum_{k=1}^{K} t_k \mod \Lambda$, and a given coset $C_i$, (45) only has one solution for $x$. Since there are $K^N$ such choices of $C_i$ and hence $K^N$ such equations as shown in (45), (43) has at most $K^N$ solutions. Hence each $\sum_{k=1}^{K} t_k \mod \Lambda$ corresponds to at most $K^N$ points of $\sum_{k=1}^{K} t_k$. This completes the proof of the theorem.

## Appendix B
## Proof of Theorem 4

### A. Some Useful Results on Decoding Nested Lattice Codes

In this section, we introduce some supporting results adapted from [25] which we will use in proving Theorem 4.

Consider $K + 1$ $N$-dimensional lattices $\Lambda_{p,i}, i = 0, ..., K$, such that $\Lambda_{p,i} \subset \mathbf{R}^N, i = 0, ..., K$ is Rogers-good for covering and Poltyrev-good for channel coding. The definition for a lattice to be Rogers-good can be found in [45, Section II.B]. The definition of Poltyrev-good can be found in [45, Section III.D]. Reference [45] gaurantees the existence of $\Lambda_{p,i}, i = 0, ..., K$ which are good in the above sense.

Construct the fine lattice $\Lambda$ as in [25, Section 7] such that $\Lambda_{p,0} \subset \Lambda$. Hence $\{\Lambda, \Lambda_{p,0}\}$ forms a nested lattice pair.

Define independent random variables $U_0^N, U_1^N, ...U_K^N$, such that $U_i^N, i = 0, ..., K$ is uniformly distributed over the fundamental region of $\Lambda_{p,i}$.

Define $\sigma^2(U_i), i = 0, ..., K$ as the variance per dimension of $U_i^N$. When $N$ increases, we scale $\Lambda$ and $\Lambda_{p,i}, i = 0, ..., K$ such that $\sigma^2(U_i)$ remains unchanged. Note that scaling does not affect the goodness of a lattice.

Define $O_i^N, i = 1, ..., K$, as zero mean Gaussian random variables such that

$$O_i^N \sim \mathcal{N}(\mathbf{0}, \sigma^2(O_i^N)\mathbf{I}) \tag{46}$$

where $\mathbf{I}$ is an $N \times N$ identity matrix. $\sigma^2(O_i^N)$ is a scalar chosen as the the variance per dimension of a random variable uniformly distributed over the smallest ball covering $\mathcal{V}(\Lambda_{p,i})$.

Define $\varepsilon(\Lambda_{p,i})$ is as in [25, (67)]:

$$\varepsilon(\Lambda_{p,i}) = \log\left(\frac{R_{u,i}}{R_{l,i}}\right) + \frac{1}{2}\log 2\pi e G_N^* + \frac{1}{N} \tag{47}$$

where $R_{u,i}, R_{l,i}$ are the covering radius and effective radius of $\Lambda_{p,i}$ respectively. $G_N^*$ is the normalized average power of $N$-dimensional sphere and converges to $\frac{1}{2\pi e}$ as $N \to \infty$. Since $\Lambda_{p,i}$ is good for covering, we have $\frac{R_{u,i}}{R_{l,i}} \to 1$ as $N \to \infty$ [25, Section 7]. Hence

$$\lim_{N \to \infty} \varepsilon(\Lambda_{p,i}) = 0 \tag{48}$$

$\sigma^2(O_i^N)$ is bounded by [25, Lemma 6]:

$$\frac{N}{N+2}\sigma^2(U_i) \leq \sigma^2(O_i^N) \leq \left(\frac{R_{u,i}}{R_{l,i}}\right)^2 \sigma^2(U_i) \tag{49}$$

Hence we have

$$\lim_{N \to \infty} \sigma^2\left(O_i^N\right) = \sigma^2(U_i) \tag{50}$$

Define $Z^N$ as a $N$-dimensional vector which is composed of zero mean i.i.d. Gaussian random variables, each with variance $\sigma^2$. $Z^N$ is independent from $U_i^N, i = 1, ..., K$.

Define $t^N$ as a lattice point in $\Lambda \cap \mathcal{V}(\Lambda_{p,0})$. $t^N$ is independent from $Z^N$ and $U_i^N, i = 1, ..., K$. Define $Y^N$ as

$$Y^N = \left(t^N + \sum_{i=1}^K U_i^N + Z^N\right) \bmod \Lambda_{p,0} \tag{51}$$

for $K \geq 1$, and

$$Y^N = \left(t^N + Z^N\right) \bmod \Lambda_{p,0} \tag{52}$$

when $K = 0$.

Define $\tilde{t}^N$ as the value for $t^N$ decoded from $Y^N$ using an Euclidean distance decoder:

$$\tilde{t}^N = \arg \min_{u^N \in \Lambda \cap \mathcal{V}(\Lambda_{p,0})} \left\| Y^N - u^N \right\|_2 \tag{53}$$

Define the rate $R_0$ of the nested lattice code book $\Lambda \cap \mathcal{V}(\Lambda_{p,0})$ as

$$R_0 = \frac{1}{N} \log_2 |\mathcal{V}(\Lambda_{p,0}) \cap \Lambda| \tag{54}$$

where $|S|$ is the cardinality of a set $S$.

With these notations, we have the following results:

*Lemma 2:* If $K = 0$ and hence $Y^N$ is given by (52), and

$$R_0 < \frac{1}{2} \log_2 \left( \frac{\sigma^2 (U_0)}{\sigma^2} \right) \tag{55}$$

Then for each $N$ dimensions there exist lattices $\Lambda, \Lambda_{p,0}$ such that

$$\Pr \left( Z^N \notin \mathcal{V}(\Lambda) \right) \tag{56}$$

decreases exponentially fast with $N$.

*Proof:* The lemma follows by repeating the proof of [25, Theorem 5] when we choose the scalar $\alpha$ defined therein to be 1 and consequently $\mathbf{N}''$ defined therein becomes $\mathbf{N}'' = (1 - \alpha)\mathbf{U} + \alpha\mathbf{N} = \mathbf{N}$. ∎

Lemma 2 can be extended into the following form:

*Lemma 3:* When $K \geq 1$, and hence $Y^N$ is given by (51), and

$$R_0 < \frac{1}{2} \log_2 \left( \frac{\sigma^2 (U_0)}{\sigma^2 + \sum\limits_{i=1}^{K} \sigma^2 (U_i)} \right) \tag{57}$$

Then for each $N$ dimensions there exist lattices $\Lambda, \Lambda_{p,i}, t = 0, ... K$ such that $\Pr(t^N \neq \tilde{t}^N)$ decreases exponentially fast with $N$.

*Proof:* Let $f_X(x)$ denote the probability density function of any continuous random variable $X$. We first use the following fact shown in [25, (200)]:

$$f_{U_i^N}(x) \leq e^{N\varepsilon(\Lambda_{p,i})} f_{O_i^N}(x) \tag{58}$$

which "approximates" $U_i^N$ with the Gaussian random variable $O_i^N$. $\varepsilon(\Lambda_{p,i})$ is defined in (47).

From (58), since $U_i^N, i = 1, ..., K, Z^N$ are independent, we have

$$f_{\sum_{i=1}^{K} U_i^N + Z^N}(x) \leq e^{N \sum_{i=1}^{K} \varepsilon(\Lambda_{p,i})} f_{\sum_{i=1}^{K} O_i^N + Z^N}(x) \tag{59}$$

which means:

$$\Pr\left(\sum_{i=1}^{K} U_i^N + Z^N \notin \mathcal{V}(\Lambda)\right) \tag{60}$$

$$\leq e^{N \sum_{i=1}^{K} \varepsilon(\Lambda_{p,i})} \Pr\left(\sum_{i=1}^{K} O_i^N + Z^N \notin \mathcal{V}(\Lambda)\right) \tag{61}$$

From Lemma 2, since $\sum_{i=1}^{K} O_i^N + Z^N$ is an i.i.d. Gaussian vector, we have, for a given $K$,

$$\Pr\left(\sum_{i=1}^{K} O_i^N + Z^N \notin \mathcal{V}(\Lambda)\right) \tag{62}$$

decreases exponentially fast with $N$ if

$$R_0 < \frac{1}{2}\log_2\left(\frac{\sigma^2(U_0)}{\sigma^2 + \lim_{N \to \infty} \sum_{i=1}^{K} \sigma^2(O_i^N)}\right) \tag{63}$$

which yields (57) because of (50). This, along with (48), means (61) decreases exponentially fast with $N$. Therefore (60) decreases exponentially fast with $N$. Since (60) is an upper bound on $\Pr(t^N \neq \tilde{t}^N)$, we have proved Lemma 3. ∎

The following result is adapted from [25, (89)].

*Lemma 4:* Define $\mu$ as

$$\mu = \frac{\sigma^2(U_0)}{\sigma^2 + \sum_{i=1}^{K} \sigma^2(U_i)} \tag{64}$$

Then if $\mu > 1$, the probability

$$\Pr\left(\sum_{i=1}^{K} U_i^N + Z^N \bmod \Lambda_{p,0} \neq \sum_{i=1}^{K} U_i^N + Z^N\right) \tag{65}$$

decreases exponentially fast with respect to $N$.

*Proof:* (65) is upper bounded by

$$\Pr\left(\sum_{i=1}^{K} U_i^N + Z^N \notin \mathcal{V}(\Lambda_{p,0})\right) \tag{66}$$

which, in turn, by following similar steps which lead to (61), is upper bounded by

$$e^{N\sum_{i=1}^{K}\varepsilon(\Lambda_{p,i})}\Pr\left(\sum_{i=1}^{K}O_i^N + Z^N \notin \mathcal{V}\left(\Lambda_{p,0}\right)\right) \tag{67}$$

Since $\Lambda_{p,0}$ is taken from the lattice code ensemble defined in [25, Section 7], according to [25, (78)], we have

$$\Pr\left(\sum_{i=1}^{K}O_i^N + Z^N \notin \mathcal{V}\left(\Lambda_{p,0}\right)\right) \le e^{-N(E_P(\mu) - o_N(1))} \tag{68}$$

where $E_P(\mu)$ is the Poltyrev exponent defined in [25, (56)]. $o_N(1)$ is any function of $N$ such that $\lim_{N\to\infty} o_N(1) = 0$ [25, Section 1]. Since $E_p(\mu)$ is positive for $\mu > 1$, we have proved Lemma 4. ∎

### B. The Coding Scheme in Theorem 4

Let $X_k^N$ be the signal sent by node $S_k$ over $N$ channel uses. $X_k^N$ is the sum of codewords from $M$ layers as shown below:

$$X_k^N = \sum_{i=1}^{M} X_{k,i}^N, \quad k = 1, 2 \tag{69}$$

$X_{k,i}^N$ is the signal sent by the $S_k$ in the $i$th layer.

For each layer, we use the nested lattice code described in Section II. Let $\{\Lambda_i, \Lambda_{c,i}\}$ be nested lattice pair assigned to layer $i$. This means $\Lambda_{c,i} \subset \Lambda_i$ and $\Lambda_{c,i}$ is Roger-good for covering and Poltyrev-good for channel coding. A construction of such a nested lattice pair can be found in [25, Section 7].

The signal $X_{k,i}^N$ is computed according to this nested lattice pair as:

$$X_{k,i}^N = \left(u_{k,i}^N + d_{k,i}^N\right) \bmod \Lambda_{c,i} \quad k = 1, 2, i = 1, ..., M \tag{70}$$

where $d_{k,i}^N$ is the dithering vector, uniformly distributed over $\mathcal{V}\left(\Lambda_{c,i}\right)$, perfectly known by all receiving nodes and independently generated for each node, each layer and each block of $N$ channel uses. Let $u_{k,i}^N$ be the lattice point such that:

$$u_{k,i}^N \in \mathcal{V}\left(\Lambda_{c,i}\right) \cap \Lambda_i, \quad k = 1, 2 \tag{71}$$

Note that both node $S_1$ and $S_2$ use the same lattice codebook for each layer.

Define $R_i$ as the rate of the codebook for the $i$th layer:

$$R_i = \frac{1}{N} \log_2 |\mathcal{V}(\Lambda_{c,i}) \cap \Lambda_i| \tag{72}$$

Since $d_{k,i}^N$ is uniformly distributed over $\mathcal{V}(\Lambda_{c,i})$, the average power per dimension of the $i$th layer $P_i$ is given by:

$$P_i = \frac{1}{N \, \mathbf{vol}(\mathcal{V}(\Lambda_{c,i}))} \int_{x \in \mathcal{V}(\Lambda_{c,i})} \|x\|_2^2 \, dx \tag{73}$$

where $\mathbf{vol}(\mathcal{V}(\Lambda_{c,i}))$ is the volume of the set $\mathcal{V}(\Lambda_{c,i})$.

As shown in (23), Node $D_1$ receives $qY_1$, which, due to (69), can be written as:

$$qY_1 = \sum_{t=1}^{M} \left( qX_{1,t}^N + (p+\gamma) X_{2,t}^N \right) + q\sqrt{b} Z_1^N \tag{74}$$

When decoding layer $i$, we assume the decoder at node $D_1$ starts from:

$$\sum_{t=1}^{i} \left( qX_{1,t}^N + (p+\gamma) X_{2,t}^N \right) + q\sqrt{b} Z_1^N \tag{75}$$

Node $D_1$ first decode $qu_{1,i}^N + pu_{2,i}^N \mod \Lambda_{c,i}$, then decode $u_{2,i}^N$.

Note that since $p$ and $q$ are both positive integers, $qu_{1,i}^N + pu_{2,i}^N \mod \Lambda_{c,i} \in \Lambda_i \cap \mathcal{V}(\Lambda_{c,i})$.

In order to decode $qu_{1,i}^N + pu_{2,i}^N \mod \Lambda_{c,i}$, the decoder computes

$$\hat{Y}_i = [\left( qu_{1,i}^N + pu_{2,i}^N \right) + \gamma X_{2,i}^N + \sum_{t=1}^{i-1} \left( qX_{1,t}^N + (p+\gamma) X_{2,t}^N \right) + q\sqrt{b} Z_1^N] \mod \Lambda_{c,i} \tag{76}$$

from (75) since it knows $d_{1,i}^N$ and $d_{2,i}^N$. Note that in (76), even though both the signal $qu_{1,i}^N + pu_{2,i}^N \mod \Lambda_{c,i}$ and the noise term $\gamma X_{2,i}^N + \sum_{t=1}^{i-1} \left( qX_{1,t}^N + (p+\gamma) X_{2,t}^N \right) + q\sqrt{b} Z_1^N$ contains $u_{2,i}^N$, they are independent because $u_{2,i}^N$ is independent from $X_{2,i}^N$ due to the dithering vector $d_{2,i}^N$. Hence Lemma 3 applies. Note that the same technique was used in [25, Lemma 2].

Define $A_i$ as

$$A_i = \sum_{t=1}^{i-1} \left( q^2 + (p+\gamma)^2 \right) P_t + q^2 b \tag{77}$$

Then, according to Lemma 3, the probability that node $D_1$ does not correctly decode $qu_{1,i}^N + pu_{2,i}^N \mod \Lambda_{c,i}$, decreases exponentially fast with the lattice dimension $N$ if

$$R_i \leq \frac{1}{2} \log_2 \left( \frac{P_i}{\gamma^2 P_i + A_i} \right) \tag{78}$$

After decoding $qu_{1,i}^N + pu_{2,i}^N \mod \Lambda_{c,i}$, node $D_1$ can recover:

$$[\gamma X_{2,i}^N + \sum_{t=1}^{i-1} \left( qX_{1,t}^N + (p+\gamma) X_{2,t}^N \right) + q\sqrt{b} Z_1^N] \mod \Lambda_{c,i} \tag{79}$$

from (76).

According to Lemma 4, as long as

$$P_i > \gamma^2 P_i + A_i \tag{80}$$

(79) equals

$$\gamma X_{2,i}^N + \sum_{t=1}^{i-1} \left( q X_{1,t}^N + (p + \gamma) X_{2,t}^N \right) + q\sqrt{b} Z_1^N \tag{81}$$

with high probability. If (79) does not equal (81), a decoding error is said to occur.

Node $D_1$ then evaluates the following expression from (81):

$$\left[ k \left( \gamma X_{2,i}^N + \sum_{t=1}^{i-1} \left( q X_{1,t}^N + (p + \gamma) X_{2,t}^N \right) + q\sqrt{b} Z_1^N \right) - \gamma d_{2,i}^N \right] \bmod \gamma \Lambda_{c,i} \tag{82}$$

$$= \left[ \gamma u_{2,i}^N + (k-1) \gamma X_{2,i}^N + k \left( \sum_{t=1}^{i-1} \left( q X_{1,t}^N + (p + \gamma) X_{2,t}^N \right) + q\sqrt{b} Z_1^N \right) \right] \bmod \gamma \Lambda_{c,i} \tag{83}$$

where the scalar $k$ corresponds to $\alpha$ in [25, (13)]. Like [25], we choose $k$ to minimize the variance per dimension of the term

$$(k-1) \gamma X_{2,i}^N + k \left( \sum_{t=1}^{i-1} \left( q X_{1,t}^N + (p + \gamma) X_{2,t}^N \right) + q\sqrt{b} Z_1^N \right) \tag{84}$$

(84) is called the "effective noise" in [25] and its minimal variance per dimension with the optimal $k$ is given by:

$$\frac{\gamma^2 P_i A_i}{\gamma^2 P_i + A_i} \tag{85}$$

We then apply Lemma 3, which says node $D_1$ can correctly decode $u_{2,i}^N$ from (83) with high probability if

$$R_i \leq \frac{1}{2} \log_2 \left( \frac{\gamma^2 P_i}{\left( \frac{\gamma^2 P_i A_i}{\gamma^2 P_i + A_i} \right)} \right) = \frac{1}{2} \log_2 \left( 1 + \frac{\gamma^2 P_i}{A_i} \right) \tag{86}$$

After decoding $u_{2,i}^N$, node $D_1$ can recover the following signal from (81):

$$\sum_{t=1}^{i-1} \left( q X_{1,t}^N + (p + \gamma) X_{2,t}^N \right) + q\sqrt{b} Z_1^N \tag{87}$$

which will be used when decoding lower layers.

After describing the decoding procedure, we proceed to determine the power $P_i$ and the rate $R_i$ of each layer. As in [38], [43], we let the right hand side of (78) equal the right hand side of (86):

$$\frac{P_i}{\gamma^2 P_i + A_i} = 1 + \frac{\gamma^2 P_i}{A_i} \tag{88}$$

It is easy to check that (88) has the following solution[2]:

$$P_i = \alpha A_i \tag{89}$$

where $\alpha = \frac{1 - 2\gamma^2 + \sqrt{1 - 4\gamma^2}}{2\gamma^4}$. This leads to (26) in Theorem 4. For $\alpha$ to be real, we require $1 - 4\gamma^2 \geq 0$, which means $|\gamma| \leq 0.5$.

By solving (89) and (77), we find that $P_i$ is given by:

$$P_i = \alpha \left(\alpha\beta + 1\right)^{i-1} q^2 b \tag{90}$$

where $\beta = q^2 + (p + \gamma)^2$. This leads to (27).

For this power allocation, $A_i$ is given by

$$A_i = \left(\alpha\beta + 1\right)^{i-1} q^2 b \tag{91}$$

Due to (88), $R_i$ can be found by averaging (78) and (86) and substituting (90) and (91) into the result:

$$R_i = \frac{0.5}{2} \log_2 \left( \frac{P_i}{\gamma^2 P_i + A_i} \right) + \frac{0.5}{2} \log_2 \left( 1 + \frac{\gamma^2 P_i}{A_i} \right) \tag{92}$$

$$= \frac{0.5}{2} \log_2 \left( \frac{P_i}{A_i} \right) \tag{93}$$

$$= 0.25 \log_2 \left( \alpha \right) \tag{94}$$

It remains to check the requirement (80). To do that, we substitute (89) into (80) and get

$$\left(1 - \gamma^2\right) \alpha > 1 \tag{95}$$

where $\alpha$ can be expressed in terms of $\gamma$ as shown in (26). It can be verified that the left hand side of (95) is always greater than 1 if $|\gamma| < 0.5$. Hence (80) is fulfilled.

The total power consumed by node $S_i$ can be computed from (90) and is given by

$$\sum_{t=1}^{M} P_t = \frac{\left(\alpha\beta + 1\right)^M - 1}{\beta} q^2 b \tag{96}$$

Having established the rate of the nested lattice codebook at each layer in (94), we next compute the secrecy rate. For this purpose, we require that $u_{2,i}^N, i = 1, ..., M$ are independent and are also independent from $u_{1,i}^N, i = 1, ..., M$. Each $u_{2,i}^N, i = 1, ..., M$ must be uniformly distributed over $\mathcal{V}(\Lambda_{c,i}) \cap \Lambda_i$. Then, if we view $u_{k,i}^N, k = 1, 2, i = 1, ..., M$ as the inputs to the channel, the signal transmitted by node $S_2$ is independent between every block of $N$-channel uses. Hence Theorem 2 applies. This means any secrecy rate $R_e$ such that

$$0 \leq R_e \leq [\lim_{N \to \infty} \frac{1}{N}(I(u_{1,i}^N, i = 1, ..., M; Y_1^N, d_{k,i}^N, k = 1, 2, i = 1, ..., M) -$$
$$I(u_{1,i}^N, i = 1, ..., M; Y_2^N, d_{k,i}^N, k = 1, 2, i = 1, ..., M))]^+ \qquad (97)$$

is achievable. Hence we only need to find a lower bound to the right hand side of (97). To do that, we first need to determine the distribution of $u_{1,i}^N, i = 1, ..., M$. Here we choose it as the same distribution we chose for $u_{2,i}^N, i = 1, ..., M$. This means $u_{1,i}^N, i = 1, ..., M$ are independent and each of them is uniformly distributed over $\mathcal{V}(\Lambda_{c,i}) \cap \Lambda_i$. For this distribution,

$$\lim_{N \to \infty} \frac{1}{N} I(u_{1,i}^N, i = 1, ..., M; Y_1^N, d_{k,i}^N, k = 1, 2, i = 1, ..., M) \qquad (98)$$

$$\leq \lim_{N \to \infty} \frac{1}{N} H(u_{1,i}^N, i = 1, ..., M) = \sum_{i=1}^M R_i \qquad (99)$$

---

[2]The other solution is when $\alpha = \frac{1 - 2\gamma^2 - \sqrt{1 - 4\gamma^2}}{2\gamma^4}$. It turns out this solution does not achieve positive secrecy rate.

On the other hand, we know that, for a given $M$, $u_{1,i}^N, i = 1, ..., M$ can be decoded from $\{Y_1^N, d_{k,i}^N, k = 1, 2, i = 1, ..., M\}$ using the decoding procedure described in this proof. By Lemma 3 and Lemma 4, the probability of decoding error decreases exponentially fast with respect to $N$. If $P_e$ denotes the probability of decoding error, then, by Fano's inequality [46], we have

$$\frac{1}{N} H\left(u_{1,i}^N, i = 1, ..., M | Y_1^N, d_{k,i}^N, k = 1, 2, i = 1, ..., M\right) \qquad (100)$$

$$\leq \frac{1}{N}\left(1 + P_e H(u_{1,i}^N, i = 1, ..., M)\right) = \frac{1}{N} + P_e \sum_{i=1}^M R_i \qquad (101)$$

Therefore

$$\lim_{N \to \infty} \frac{1}{N} I(u_{1,i}^N, i = 1, ..., M; Y_1^N, d_{k,i}^N, k = 1, 2, i = 1, ..., M) \qquad (102)$$

$$= \lim_{N \to \infty} \frac{1}{N}(H(u_{1,i}^N, i = 1, ..., M) - H\left(u_{1,i}^N, i = 1, ..., M | Y_1^N, d_{k,i}^N, k = 1, 2, i = 1, ..., M\right)) \qquad (103)$$

$$\geq \sum_{i=1}^M R_i - \frac{1}{N} - P_e \sum_{i=1}^M R_i \qquad (104)$$

By letting $N \to \infty$, (99) and (104) imply:

$$\lim_{N \to \infty} \frac{1}{N}(I(u_{1,i}^N, i = 1, ..., M; Y_1^N, d_{k,i}^N, k = 1, 2, i = 1, ..., M) = \sum_{i=1}^M R_i \qquad (105)$$

where $R_i$ is given by (94).

The second term in (97) can be upper bounded as follows:

$$\frac{1}{N} I\left(u_{1,i}^N, i = 1...M; Y_2^N, d_{k,i}^N, k = 1, 2, i = 1...M\right) \tag{106}$$

$$\leq \frac{1}{N} I\left(u_{1,i}^N, i = 1...M; X_1^N \pm X_2^N, d_{k,i}^N, k = 1, 2, i = 1...M\right) \tag{107}$$

$$\leq \frac{1}{N} \sum_{i=1}^M I\left(u_{1,i}^N; X_{1,i}^N \pm X_{2,i}^N, d_{k,i}^N, k = 1, 2\right) \tag{108}$$

$$\leq M \tag{109}$$

(107) follows from the fact that telling the eavesdropper the channel noise $Z_2^N$ will not harm its ability to obtain more information about the confidential message. (108) is because the jamming signal $X_{2,i}^N$ and the dithering noise $d_{k,i}^N$ of different layers are independent from each other. Finally, we apply (5)-(10) to each term inside the sum in (108) to obtain (109).

Substituting (105) and (109) into (97), we find that the following secrecy rate is achievable.

$$R_e = [\sum_{i=1}^M R_i - M]^+ \tag{110}$$

$$= [(0.25 \log_2(\alpha) - 1)M]^+ \tag{111}$$

The secure degree of freedom is therefore given by

$$\lim_{P \to \infty} \frac{R_e}{\frac{1}{2} \log_2 P} = \lim_{M \to \infty} \frac{[\sum_{i=1}^M R_i - M]^+}{\frac{1}{2} \log_2 \sum_{t=1}^M P_t} \tag{112}$$

$$= \left[\frac{0.25 \log_2(\alpha) - 1}{\frac{1}{2} \log_2(\alpha\beta + 1)}\right]^+ \tag{113}$$

Hence, we have proved Theorem 4.

*Remark 15:* $d_{k,i}^N$ can be replaced with deterministic vectors. The proof is given in Appendix C. □

*Remark 16:* We scaled the signal by $k$ in (83) before performing the modulus operation. Doing so offers a slight gain in secure degree of freedom than just choosing $k = 1$.

The scaling by $k$ operation can also be done in (76). However, the optimal scaling factor has a more complicated expression and it is difficult to derive an analytical expression for the secure degree of freedom if this approach is followed. □

## Appendix C

### Replacing Random Dithering Vectors with Deterministic Vectors

In this appendix, we shall demonstrate that the dithering noise $d_{i,k}^N$ in Appendix B can be replaced with deterministic vectors. For a fix number of layers $M$, let $e_i$ denotes the binary random variable such that $e_i = 1$ denote the event that a decoding error has occurred at layer $i$ given no decoding error has occurred at upper layers $j$, $M \leq j > i$. There are three types of error events that results in $e_i = 1$, depending on:

1)  whether node $D_1$ can decode $qu_{1,i}^N + pu_{2,i}^N \mod \Lambda_{c,i}$ from $\hat{Y}_i$ given by (76) correctly.

2)  whether (79) equals (81).

3)  whether node $D_1$ can decode $u_{2,i}^N$ from (83) correctly.

By Lemma 3 and Lemma 4, the probability of each error event goes to $0$ exponentially fast as the dimension of the lattice $N$ goes to $\infty$. Hence from the union bound, we have:

$$\Pr(e_i) < \exp(-N\tilde{\alpha}_i) \tag{114}$$

where $\tilde{\alpha}_i$ is the error exponent of the error event $\{e_i = 1\}$. Let $\tilde{\alpha} = \min_i \tilde{\alpha}_i$. Let $e$ be the binary random variable such that $e = 1$ means a decoding error has occurred. Then $\{e = 1\} = \cup_{i=0}^{M-1}\{e_i = 1\}$. From the union bound, we have:

$$\Pr(e) < M \exp(-N\tilde{\alpha}) \tag{115}$$

Another way to view this result, as described in [35], is that (115) is the average performance of an ensemble of codebooks. Each codebook is denoted by $\mathcal{C}_k, k = 1, 2$ with parameter $d_{k,i}^N, i = 1, ..., M, k = 1, 2$, where

$$\mathcal{C}_k = \{ (u_{k,i}^N + d_{k,i}^N) \mod \Lambda_{c,i} : u_{k,i}^N \in \mathcal{V}(\Lambda_{c,i}) \cap \Lambda_i, i = 1, ..., M \} \tag{116}$$

Each term inside the bracket on the right hand of (116), which is $(u_{k,i}^N + d_{k,i}^N) \mod \Lambda_{c,i}$, denotes the codebook used for each layer. Although the transmitted signal is the sum of the codewords from the codebook of each layer, they can be differentiated by the receiver due to a careful power and rate allocation between layers as shown in Appendix B.

Note that because of the nested structure, $(u_{k,i}^N + d_{k,i}^N) \mod \Lambda_{c,i}$ contains the same number of lattice points regardless of the choice of $d_{k,i}^N$ [35]. Therefore, $\mathcal{C}_k$ has same rate regardless of the choice of $d_{k,i}^N$. That said, this choice may lead to different average power and error probability performance. (115) can then be interpreted as the error probability averaged over the code ensembles and can be expressed as

$$E_{\mathcal{C}_1,\mathcal{C}_2}[\Pr(e)] < M \exp(-N\tilde{\alpha}) \tag{117}$$

where $E_{\mathcal{C}_1,\mathcal{C}_2}[\ ]$ means the expectation considering the codebook $\mathcal{C}_1, \mathcal{C}_2$ as being random. From Markov inequality, at least $(1 - \varepsilon)$ fraction of the codebook $\mathcal{C}_1$ has the following property:

$$E_{\mathcal{C}_2}[\Pr(e)|\mathcal{C}_1 = \mathcal{C}_1^*] < \frac{1}{\varepsilon} E_{\mathcal{C}_1,\mathcal{C}_2}[\Pr(e)] \tag{118}$$

Let $P_i'$ be the average power of a certain codebook $\mathcal{C}_i$ and $\bar{P}_i'$ be the average power of the codebook ensemble. At least $2\varepsilon$ fraction of the codebook $\mathcal{C}_1$ has the following property:

$$P_1' \leq \frac{1}{1 - 2\varepsilon} \bar{P}_1' \tag{119}$$

Therefore, there must exists one codebook $\mathcal{C}_1^*$ that meets (118) and (119) simultaneously. Let $\mathcal{C}_1 = \mathcal{C}_1^*$ to be this codebook, and apply the same argument above to $\mathcal{C}_2$. Then there must exists one codebook $\mathcal{C}_2^*$ such that

$$\Pr(e|\mathcal{C}_1 = \mathcal{C}_1^*, \mathcal{C}_2 = \mathcal{C}_2^*) < \frac{1}{\varepsilon} E_{\mathcal{C}_2}[\Pr(e)|\mathcal{C}_1 = \mathcal{C}_1^*] \tag{120}$$

$$P_2' \leq \frac{1}{1 - 2\varepsilon} \bar{P}_2' \tag{121}$$

Note that this is a consequence of the fact that $\mathcal{C}_1$ is independent from $\mathcal{C}_2$. Therefore conditioning on $\mathcal{C}_1^*$ does not change the distribution of $\mathcal{C}_2$. Hence conditioning on $\mathcal{C}_1^*$, the average power of $\mathcal{C}_2$ ensemble remains to be $\bar{P}_2'$.

Now choose $\varepsilon$ as a function of $N$ such that

$$\lim_{N \to \infty} \frac{1}{\varepsilon^2} E_{\mathcal{C}_1,\mathcal{C}_2}[\Pr(e)] = 0 \tag{122}$$

$$\lim_{N \to \infty} \varepsilon = 0 \tag{123}$$

which is possible by letting $\varepsilon = 1/N$, because for a given $M$, $E_{\mathcal{C}_1,\mathcal{C}_2}[\Pr(e)]$ decreases exponentially fast with $N$ as shown by (117). From (118) and (120), we know for each choice $\varepsilon$, there exists a pair of codeword $\mathcal{C}_1^*, \mathcal{C}_2^*$ such that

$$\Pr(e|\mathcal{C}_1 = \mathcal{C}_1^*, \mathcal{C}_2 = \mathcal{C}_2^*) < \frac{1}{\varepsilon^2} E_{\mathcal{C}_1,\mathcal{C}_2}[\Pr(e)] \tag{124}$$

and the power consumption meets the (119) and (121). This, along with (122) and (123), means that there must exists a pair of codebooks $(\mathcal{C}_1 = \mathcal{C}_1^*, \mathcal{C}_2 = \mathcal{C}_2^*)$ that meet both the error probability requirement and the power constraint requirement when $N$ goes to $\infty$. This means that the dithering vector $\{d_{i,k}^N\}$ in this work can be deterministic vectors.

## Appendix D
### Proof of Theorem 5

In this appendix we provide the proof for Theorem 5. We use the same integer lattice codebook as in [23, Theorem 1]. Let $\Lambda_{P,\varepsilon}$ be the scalar lattice defined as:

$$\Lambda_{P,\varepsilon} = \left\{ x : x = P^{1/4+\varepsilon}z, z \in \mathbf{Z} \right\} \tag{125}$$

The codebook $\mathcal{C}_{P,\varepsilon}$ is given by:

$$\mathcal{C}_{P,\varepsilon} = \Lambda_{P,\varepsilon} \cap \left[ -\sqrt{P}, \sqrt{P} \right] \tag{126}$$

where $P = \min\{\bar{P}_1, \bar{P}_2\}$. Hence $\left\lceil -P^{1/4-\varepsilon} \right\rceil \leq z \leq \left\lfloor P^{1/4-\varepsilon} \right\rfloor$, where $\lceil\ \rceil$ and $\lfloor\ \rfloor$ denote the up and down rounding operations respectively. $|\mathcal{C}_{P,\varepsilon}| \geq 2\left(P^{1/4-\varepsilon}-1\right)+1 = 2P^{1/4-\varepsilon}-1$. This means that, for large enough $P$, we can write:

$$\log_2 |\mathcal{C}_{P,\varepsilon}| \geq \log_2 \left(2P^{1/4-\varepsilon}-1\right) \geq \log_2 \left(P^{1/4-\varepsilon}\right) \tag{127}$$

The same codebook is used by both nodes $S_1$ and $S_2$. The codeword transmitted by node $S_1$ is chosen based on the secret message $W_1$. The codeword transmitted by node $S_2$ is chosen independently according to a uniform distribution over $\mathcal{C}_{P,\varepsilon}$.

Since the input $X_2$ from $S_2$ is i.i.d., the channel can then be shown to be equivalent to a memoryless wiretap channel [3]. Similar to the proof of Theorem 4, we use Theorem 2 with $n = 1$, which states any secrecy rate $R_e$ such that

$$0 \leq R_e \leq \left[ I\left(X_1; Y_1\right) - I\left(X_1; Y_2\right) \right]^+ \tag{128}$$

is achievable. Hence to compute the achievable secrecy rate, we need to find a lower bound to the right hand side of (128).

According to [23, Theorem 1], $p(X_1)$ is chosen to be a uniform distribution over $\mathcal{C}_{P,\varepsilon}$. Therefore

$$H\left(X_1\right) = \log_2 \left(|\mathcal{C}_{P,\varepsilon}|\right) \tag{129}$$

From [23, Theorem 1], when

$$P > \frac{1}{a^2 b^2} \tag{130}$$

we have

$$H\left(X_1|Y_1\right) \leq 1 + 2\exp\left(-\frac{P^{2\varepsilon}}{8b}\right)\log_2\left(|\mathcal{C}_{P,\varepsilon}|\right) \tag{131}$$

The fact that $\sqrt{ab}$ is algebraic irrational is used in [23, Theorem 1], which uses a result from Diophantine approximation. (130) comes from [23, Lemma 2]. $b$ in (131) comes from the fact that the variance of the Gaussian noise contained in $Y_1$ is $b$ instead of unity.

From (129) and (131), we have:

$$I\left(X_1;Y_1\right) \geq \left(1 - 2\exp\left(-\frac{P^{2\varepsilon}}{8b}\right)\right)\log_2\left(|\mathcal{C}_{P,\varepsilon}|\right) - 1 \tag{132}$$

Since $\mathcal{C}_{P,\varepsilon}$ is simply a scaled and shifted version of the integer lattice code, we can use Lemma 1 to bound $I\left(X_1;Y_2\right)$:

$$I\left(X_1;Y_2\right) \leq I\left(X_1;Y_2,Z_2\right) = I\left(X_1;X_1 \pm X_2\right) = f(Q) \leq 0.8 \tag{133}$$

where $f(Q)$ is defined in (28). $Q$ is determined by the range of $z$ in (125). Hence $Q = 2\lfloor P^{1/4-\varepsilon}\rfloor + 1$. The last inequality in (133) follows from Lemma 1. Using (132) (133), and (127), we find (128) is lower bounded by

$$\left[\left(1 - 2\exp\left(-\frac{P^{2\varepsilon}}{8b}\right)\right)\left(\frac{1}{4}-\varepsilon\right)\log_2\left(P\right) - 1 - f(Q)\right]^+ \tag{134}$$

for sufficiently large $P$. Since $0 \leq f(Q) < 0.8$, and $\varepsilon$ can be any value between $(0, 1/4)$, using the definition in (21), we find the achieved secure degree of freedom is $0.5$. Hence we have completed the proof.

## Appendix E

### Proof of Theorem 6

In this appendix we provide the proof for Theorem 6. We begin by considering the case when $\sqrt{ab} \geq 2$. Let

$$X_k = \sqrt{P_0}\sum_{i=0}^{M-1} a_{k,i}Q^{2i}, k = 1, 2 \tag{135}$$

where $M$ is a constant positive integer, $P_0$ is a constant scaling factor. Both are related to the variance of $X_k$. $a_{k,i}$ is uniformly distributed over $[0, \lfloor Q\rfloor - 1]\cap \mathbf{Z}$. Due to the range limit imposed on $a_{k,i}$, we observe that $a_{k,i}$ is uniquely determined by $X_k$.

The signal received by node $D_1$ is given by

$$Y_1 = \sqrt{P_0}\left(\sum_{i=0}^{M-1} a_{1,i}Q^{2i} + \sum_{i=0}^{M-1} a_{2,i}Q^{2i+1}\right) + \sqrt{b}Z_1 \tag{136}$$

We then derive a lower bound to $[I(X_1; Y_1) - I(X_2; Y_2)]^+$ as we did for Theorem 5.

$I(X_1; Y_1)$ can be lower bounded by following a similar derivation from [23, Theorem 1]. Define $\mathcal{C}_k = \{\sqrt{P_0}\sum_{i=0}^{M-1} a_{k,i}Q^{2i} : a_{k,i} \in [0, \lfloor Q \rfloor -1] \cap \mathbf{Z}\}$, $k = 1, 2$. We use the same maximum likelihood decoder used in [23, Theorem 1]:

$$\hat{Y}_1 = \arg \min_{X_1 + QX_2, \text{ s.t. } X_k \in \mathcal{C}_k, k=1,2} |Y_1 - (X_1 + QX_2)|^2 \tag{137}$$

It is clear that given $\hat{Y}_1$, there is a unique pair of $X_1, X_2$ such that $X_1 + QX_2 = \hat{Y}_1$. Let this mapping from $\hat{Y}_1$ to $X_1$ be $f$. Define binary random variable $A$ such that

$$A = \begin{cases} 0 & \text{if } |\sqrt{b}Z_1| < \sqrt{P_0}/2 \\ 1 & \text{otherwise} \end{cases} \tag{138}$$

Note that if $A = 0$, we have $X_1 = f(\hat{Y}_1)$. For this definition of $A$, we have

$$H\left(X_1|\hat{Y}_1\right) \leq H\left(X_1, A|\hat{Y}_1\right) \tag{139}$$

$$= H\left(A|\hat{Y}_1\right) + H\left(X_1|\hat{Y}_1, A\right) \tag{140}$$

$$\leq 1 + \Pr\left(A = 1\right) H\left(X_1|\hat{Y}_1, A = 1\right) + \Pr\left(A = 0\right) H\left(X_1|\hat{Y}_1, A = 0\right) \tag{141}$$

$$= 1 + \Pr\left(A = 1\right) H\left(X_1|\hat{Y}_1, A = 1\right) \tag{142}$$

$$\leq 1 + \Pr\left(A = 1\right) H\left(X_1|A = 1\right) \tag{143}$$

$$= 1 + \Pr\left(A = 1\right) H\left(X_1\right) \tag{144}$$

(144) is because $A$ is independent from $X_1$. $\Pr(A = 1)$ is bounded as follows:

$$\Pr\left(A = 1\right) = \int_{|t| \geq \sqrt{P_0}/2} \frac{1}{\sqrt{2\pi b}} \exp\left(-\frac{t^2}{2b}\right) dt \leq 2 \exp\left(-\frac{P_0}{8b}\right) \tag{145}$$

Substituting it into (144), we get:

$$H\left(X_1|\hat{Y}_1\right) \leq 1 + 2 \exp\left(-\frac{P_0}{8b}\right) H\left(X_1\right) \tag{146}$$

Therefore $I(X_1; Y_1)$ is lower bounded as:

$$I\left(X_1; Y_1\right) \geq I\left(X_1; \hat{Y}_1\right) \geq \left(1 - 2 \exp\left(-\frac{P_0}{8b}\right)\right) H\left(X_1\right) - 1 \tag{147}$$

For $I(X_1; Y_2)$, we have:

$$I(X_1; Y_2) \leq I(X_1; X_1 \pm X_2) \leq \sum_{i=0}^{M-1} I(a_{1,i}; a_{1,i} \pm a_{2,i}) = Mf(\lfloor Q \rfloor) \tag{148}$$

Given $X_1$ as defined in (135), we have $H(X_1) = M\log_2\lfloor Q \rfloor$. Substituting it into (147) and combining it with (148), we find, from (128), the following secrecy rate is achievable.

$$R_e = [M(1 - 2\exp(-\frac{P_0}{8b}))(\log_2\lfloor Q \rfloor) - 1 - Mf(\lfloor Q \rfloor)]^+ \tag{149}$$

The transmission power is given by:[3]

$$Var\,[X_k] = P_0\left(\frac{\lfloor Q \rfloor^2 - 1}{12}\right)\sum_{i=0}^{M-1} Q^{4i} \tag{150}$$

$$= P_0\left(\frac{\lfloor Q \rfloor^2 - 1}{12}\right)\frac{Q^{4M} - 1}{Q^4 - 1} \tag{151}$$

The secure degree of freedom can then be computed by substituting the transmission power (151) and the secrecy rate (149) into (21), which yields:

$$\lim_{M\to\infty} \frac{\left[\left(\left(1 - 2\exp\left(-\frac{P_0}{8b}\right)\right)\log_2\lfloor Q \rfloor - f(\lfloor Q \rfloor)\right)M\right]^+}{\frac{1}{2}\log_2\left(Q^{4M}\right)} \tag{152}$$

$$= [\frac{1}{2}\left(1 - 2\exp\left(-\frac{P_0}{8b}\right)\right)\frac{\log_2\lfloor Q \rfloor}{\log_2 Q} - \frac{f(\lfloor Q \rfloor)}{2\log_2(Q)}]^+ \tag{153}$$

(153) can be made arbitrarily close to (32) by choosing a large enough $P_0$. (33) then follows from (32) via Lemma 1.

When $Q = 2$, it can be verified that $f(\lfloor Q \rfloor) = 0.5$, and (153) can be made to be arbitrarily close to $1/4$.

The case of $1/\sqrt{ab} \geq 2$ can be proved in a similar fashion. Let $1/\sqrt{ab} = Q$ and let $X_k = Q\sqrt{P_0}\sum_{i=0}^{M-1} a_{k,i}Q^{2i}$, $k = 1, 2$. Then all previous derivations apply. In particular, (136) becomes:

$$Y_1 = \sqrt{P_0}\left(\sum_{i=0}^{M-1} a_{1,i}Q^{2i+1} + \sum_{i=0}^{M-1} a_{2,i}Q^{2i}\right) + \sqrt{b}Z_1 \tag{154}$$

and (137) becomes:

$$\hat{Y}_1 = \arg\min_{X_1 + X_2/Q, \text{ s.t. } X_k \in \mathcal{C}_k, k=1,2} |Y_1 - (X_1 + \frac{1}{Q}X_2)|^2 \tag{155}$$

---

[3]In case it is desired for $X_k$ to have zero mean, we can simply shift $X_k$ by a constant, which will not alter the secrecy rate.

The achieved secrecy rate remains the same. The transmission power is scaled by $Q^2$:

$$Var\left[X_k\right] = Q^2 P_0 \left(\frac{\lfloor Q \rfloor^2 - 1}{12}\right)\frac{Q^{4M} - 1}{Q^4 - 1}, \quad k = 1, 2 \tag{156}$$

Hence the secure degree of freedom is still given by (33) when $Q > 2$ and $0.25$ when $Q = 2$.

## Appendix F
### Proof of Theorem 7

Let $X_k$ be given by (135) in Section E with $Q = 2$. However, unlike what we did in (135), $a_{k,i}$ is not uniformly distributed over $\{0, 1\}$. Instead, we choose its distribution to maximize

$$I(a_{1,i}; a_{1,i} + a_{2,i}) - I(a_{1,i}; a_{1,i} - a_{2,i}) \tag{157}$$

which is about $0.1095$ when $\Pr(a_{1,i} = 1) = 0.1443$, $\Pr(a_{2,i} = 1) = 0.8557$. We next derive the achievable secrecy rate by deriving a lower bound on $[I(X_1; Y_1) - I(X_2; Y_2)]^+$. Define $\mathcal{C}_k = \{\sqrt{P_0}\sum_{i=0}^{M-1} a_{k,i}Q^{2i} : a_{k,i} \in \{0, 1\}\}$, $k = 1, 2$. Define $f(Y_1)$ as

$$f(Y_1) = \arg\min_{X_1 + X_2 : X_k \in \mathcal{C}_k, k = 1, 2} |Y_1 - (X_1 + X_2)|^2 \tag{158}$$

Then:

$$I\left(X_1; X_1 + X_2\right) - I\left(X_1; f\left(Y_1\right)\right)$$

$$\leq I\left(X_1; X_1 + X_2 | f\left(Y_1\right)\right) \tag{159}$$

$$\leq H\left(X_1 + X_2 | f\left(Y_1\right)\right) \tag{160}$$

$$\leq 1 + 2\exp\left(-\frac{P_0}{8b}\right) H\left(X_1 + X_2\right) \tag{161}$$

Inequality (161) follows from (139)-(144) with $\hat{Y}_1$ replaced by $f(Y_1)$ defined in (158), and $X_1$ replaced with $X_1 + X_2$. Then we have

$$I\left(X_1; Y_1\right)$$

$$\geq I\left(X_1; f\left(Y_1\right)\right) \tag{162}$$

$$\geq \left(1 - \left(2\exp\left(-\frac{P_0}{8b}\right)\right)\right) H\left(X_1 + X_2\right) - H\left(X_2\right) - 1 \tag{163}$$

$$= I\left(X_1; X_1 + X_2\right) - 1 - 2\exp\left(-\frac{P_0}{8b}\right) H\left(X_1 + X_2\right) \tag{164}$$

$$= \sum_{i=0}^{M-1} I\left(a_{1,i}; a_{1,i} + a_{2,i}\right) - 1 - 2\exp\left(-\frac{P_0}{8b}\right)\sum_{i=0}^{M-1} H\left(a_{1,i} + a_{2,i}\right) \tag{165}$$

In (163) we use the fact that $X_1$ is independent from $X_2$ and apply the result from (161). (165) is because there is a one-to-one mapping between $X_1 + X_2$ and $\{a_{1,i} + a_{2,i}, i = 0, ..., M - 1\}$.

For $I(X_1; Y_2)$, we have

$$I(X_1; Y_2) \leq I(X_1; X_1 - X_2) \tag{166}$$

$$\leq \sum_{i=1}^{M} I(a_{1,i}; a_{1,i} - a_{2,i}) \tag{167}$$

Therefore

$$I(X_1; Y_1) - I(X_1; Y_2)$$

$$\geq \sum_{i=1}^{M} \left( I(a_{1,i}; a_{1,i} + a_{2,i}) - I(a_{1,i}; a_{1,i} - a_{2,i}) \right) - 1 - 2\exp\left(-\frac{P_0}{8b}\right) \sum_{i=0}^{M-1} H(a_{1,i} + a_{2,i}) \tag{168}$$

$$= M0.1095 - 1 - 2\exp\left(-\frac{P_0}{8b}\right) MH(a_{1,i} + a_{2,i}) \tag{169}$$

Let $V_{a,k}$ be the variance of $a_{k,i}$. Then we have

$$Var[X_k] = P_0 V_{a,k} \frac{Q^{4M} - 1}{Q^2 - 1}, \quad k = 1, 2 \tag{170}$$

Hence the secure degree of freedom is given by:

$$\lim_{M \to \infty} \frac{[I(X_1; Y_1) - I(X_1; Y_2)]^+}{\frac{1}{2}\log_2 Var[X_k]} \tag{171}$$

$$= [0.0548 - \exp\left(-\frac{P_0}{8b}\right) H(a_{1,i} + a_{2,i})]^+ \tag{172}$$

We can always choose a large enough $P_0$ to make the secure degree of freedom to be arbitrarily close to $0.0548$. This completes the proof of Theorem 7.

## APPENDIX G

### PROOF OF THEOREM 8

Let $X_i$ be:

$$X_k = \sqrt{P_0} \sum_{i=0}^{3M-1} a_{k,i} Q^i, k = 1, 2 \tag{173}$$

where $Q = 2$. $M$ is a positive integer and $P_0$ is a positive constant as defined in the proof of Theorem 6. Choose $a_{k,i}$ such that

$$a_{k,i} = 0, \text{ if } i \bmod 3 = 1 \text{ or } 2 \tag{174}$$

otherwise $a_{k,i}$ is uniformly distributed over $\{0, 1\}$. Here $a_{k,i}$ is forced to be zero at $i \bmod 3 = 1$ or $2$ to make room for $0.5a_{2,j}, \forall j \bmod 3 = 0$ and the carry-over from $a_{1,j} + a_{2,j}, \forall j \bmod 3 = 0$.

Define $\mathcal{C}_k$ as the collection of points $X_k$ defined by (173). Define $\hat{Y}_1$ as

$$\hat{Y}_1 = \arg \min_{X_1 + 1.5X_2, \text{ s.t. } X_k \in \mathcal{C}_k, k=1,2} |Y_1 - (X_1 + 1.5X_2)|^2 \tag{175}$$

With this new definition of $\hat{Y}_1$, the same derivation that leads to (147) applies, where $H(X_1) = M$. On the other hand:

$$I\left(X_1; Y_2\right) \leq I\left(X_1; X_1 \pm X_2\right) \leq \sum_{i=0}^{3M-1} I\left(a_{1,i}; a_{1,i} \pm a_{2,i}\right) = 0.5M \tag{176}$$

From (147), (176) and (128), the following secrecy rate is achievable:

$$\left[\left(1 - 2\exp\left(-\frac{P_0}{8b}\right)\right) M - 1 - 0.5M\right]^+ \tag{177}$$

The transmission power is given by

$$Var\left[X_k\right] = P_0 \frac{Q^2 - 1}{12} \sum_{i=0}^{M-1} Q^{6i} = \frac{P_0}{4} \frac{Q^{6M} - 1}{Q^6 - 1} \tag{178}$$

for $Q = 2$. The secure degree of freedom is hence given by:

$$\lim_{M \to \infty} \frac{\left(1 - 2\exp\left(-\frac{P_0}{8b}\right)\right) M - 1 - 0.5M}{\frac{1}{2}\log_2\left(Q^{6M}\right)} \tag{179}$$

$$= \frac{1}{3}\left(1 - 2\exp\left(-\frac{P_0}{8b}\right) - 0.5\right) \tag{180}$$

which can be made arbitrarily close to $1/6$ by choosing a large enough $P_0$.

## Appendix H

### Proof of Theorem 9

We prove Theorem 9 by showing that the loss in secrecy rate due to the imperfectness of channel state information can be bounded by a constant.

The challenge in proving this result arises due to the fact that the assumption on $\alpha_i$ essentially makes the channel in Figure 10 an *arbitrary* wiretap channel for which an explicit calculation of the secrecy rate is difficult [47]. To work around this problem, we consider the channel in Figure 12 and present the following Lemma.

*Lemma 5:* Any secrecy rate achievable in Figure 12 is also achievable in Figure 10 with the *same* coding scheme.

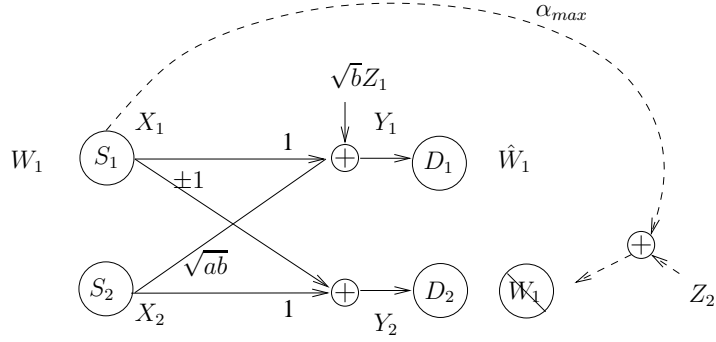

Fig. 12.   Figure 10 with an enhanced eavesdropper channel

*Proof:* We need to show the confidential message $W_1$ is reliably received by $D_1$ and the secrecy constraint (14) is met. The first requirement is met automatically since that the signal received by $D_1$ remains the same. Hence we can start with a coding scheme that reliably transmits $W_1$ from $S_1$ to $D_1$ for Figure 12 and prove that the secrecy constraint (14) is fulfilled with this coding scheme for Figure 10. For the coding scheme, the layered nested lattice codes from Theorem 4 is used.

Let $n$ be the total number of channel uses. Let $D$ be the $n \times n$ diagonal matrix, whose diagonal element at the $i$th row and $i$th column is $\alpha_i$. Define a diagonal matrix $\bar{D}$ such that $\bar{D}$ is obtained from $D$ by replacing the $0$ elements on its diagonal line with $\alpha_{\max}$. Define $Z_3^n$ as independent random variable with the same Gaussian distribution as $Z_2^n$. $Z_i^n, i = 2, 3$ are independent. Then the mutual information between $W_1$ and the knowledge of the eavesdropper in Figure 10 can be upper bounded as follows:

$$0 \leq I\left(W_1; Y_2^n, d_{k,i}^n, k = 1, 2, i = 1...M, D\right) \tag{181}$$

$$\leq I\left(W_1; X_1^n \pm X_2^n + Z_2^n + D X_1^n, d_{k,i}^n, k = 1, 2, i = 1...M, D\right) \tag{182}$$

$$= I\left(W_1; X_1^n \pm X_2^n + Z_2^n + D X_1^n, D, d_{k,i}^n, k = 1, 2, i = 1...M\right) \tag{183}$$

$$\leq I\left(W_1; X_1^n \pm X_2^n, Z_2^n + D X_1^n, D, d_{k,i}^n, k = 1, 2, i = 1...M\right) \tag{184}$$

$$\leq I\left(W_1; X_1^n \pm X_2^n, Z_2^n + \bar{D} X_1^n, \bar{D}, d_{k,i}^n, k = 1, 2, i = 1...M\right) \tag{185}$$

$$= I\left(W_1; X_1^n \pm X_2^n, \frac{\bar{D}}{\alpha_{\max}} Z_2^n + \sqrt{1 - \frac{\bar{D}^2}{\alpha_{\max}^2}} Z_3^n + \bar{D} X_1^n, \bar{D}, d_{k,i}^n, k = 1, 2, i = 1...M\right) \tag{186}$$

$$\leq I\left(W_1; X_1^n \pm X_2^n, \frac{\bar{D}}{\alpha_{\max}} Z_2^n + \bar{D}X_1^n, \bar{D}, d_{k,i}^n, k = 1, 2, i = 1...M\right) \tag{187}$$

$$= I\left(W_1; X_1^n \pm X_2^n, Z_2^n + \alpha_{\max} X_1^n, \bar{D}, d_{k,i}^n, k = 1, 2, i = 1...M\right) \tag{188}$$

Note that (188) is exactly the same as the mutual information between $W_1$ and the eavesdropper's knowledge in Figure 12. Hence if

$$\lim_{n\to\infty} \frac{1}{n} I\left(W_1; X_1^n \pm X_2^n, Z_2^n + \alpha_{\max} X_1^n, \bar{D}, d_{k,i}^n, k = 1, 2, i = 1...M\right) = 0 \tag{189}$$

Then we must have:

$$\lim_{n\to\infty} \frac{1}{n} I\left(W_1; Y_2^N, d_{k,i}^N, k = 1, 2, i = 1...M, D\right) = 0 \tag{190}$$

This means that to obtain the secrecy rate for Figure 10, we can as well compute the secrecy rate for Figure 12. ∎

Note that with the layered nested lattice coding scheme in Theorem 4, when each lattice point is viewed as a single channel use, and the channel in Figure 12 is equivalent to a memoryless wiretap channel. Then according to Theorem 2, the following secrecy rate $R_e$ is achievable:

$$0 \leq R_e \leq [\lim_{N\to\infty} \frac{1}{N}(I(u_{1,i}^N, i = 1, ..., M; Y_1^N, d_{k,i}^N, k = 1, 2, i = 1, ..., M) -$$

$$I(u_{1,i}^N, i = 1, ..., M; X_1^N \pm X_2^N, \alpha_{\max} X_1^N + Z_2^N, d_{k,i}^N, k = 1, 2, i = 1, ..., M))]^+ \tag{191}$$

The first term in (191) is still given by (105) since the signal received by $D_1$ remains the same. The second term in (191) can be upper bounded as follows:

$$\frac{1}{N} I\left(u_{1,i}^N, i = 1...M; X_1^N \pm X_2^N, Z_2^N + \alpha_{\max} X_1^N, d_{k,i}^N, k = 1, 2, i = 1...M\right) \tag{192}$$

$$= \frac{1}{N} I\left(u_{1,i}^N, i = 1...M; X_1^N \pm X_2^N, Z_2^N + \alpha_{\max} X_1^N | d_{k,i}^N, k = 1, 2, i = 1...M\right) \tag{193}$$

$$\leq \frac{1}{N} I\left(u_{1,i}^N, i = 1...M; X_1^N \pm X_2^N | d_{k,i}^N, k = 1, 2, i = 1...M\right)$$

$$+ \frac{1}{N} I\left(u_{1,i}^N, i = 1...M, X_1^N \pm X_2^N; Z_2^N + \alpha_{\max} X_1^N | d_{k,i}^N, k = 1, 2, i = 1...M\right) \tag{195}$$

$$= \frac{1}{N} I\left(u_{1,i}^N, i = 1...M; X_1^N \pm X_2^N | d_{k,i}^N, k = 1, 2, i = 1...M\right)$$

$$+ \frac{1}{N} I\left(u_{1,i}^N, i = 1...M; Z_2^N + \alpha_{\max} X_1^N | d_{k,i}^N, k = 1, 2, i = 1...M\right) \tag{196}$$

(196) is because $\{X_1^N \pm X_2^N\} - \{u_{1,i}^N, i = 1...M, d_{k,i}^N, k = 1, 2, i = 1...M\} - \{Z_2^N + \alpha_{\max}X_1^N\}$ is a Markov chain. The first term in (196) is shown by (107)-(109) to be bounded by $M$. Hence we only need to bound the second term. This term is bounded by:

$$\frac{1}{N}I\left(u_{1,i}^N, d_{k,i}^N, k = 1, 2, i = 1, ..., M; Z_2^N + \alpha_{\max}X_1^N\right) \tag{197}$$

$$\leq \frac{1}{N}I\left(X_1^N; Z_2^N + \alpha_{\max}X_1^N\right) \tag{198}$$

$$\leq C\left(\alpha_{\max}^2 P_1\right) \tag{199}$$

where $C(x) = \frac{1}{2}\log_2(1+x)$ is the channel capacity formula of average power constrained AWGN channel. $P_1$, the total transmission power consumed by $S_1$, is given by (96). The inequality (199) is due to the fact that for additive channel with i.i.d. Gaussian noise the mutual information is maximized by Gaussian input distribution for given transmitter power constraint.

Applying (199), (107)-(109), and (105) in (191), we find the secrecy rate is given by

$$R_e = [(0.25\log_2(\alpha) - 1)M - C\left(\alpha_{\max}^2 P_1\right)]^+ \tag{200}$$

Compared it to (111), the loss in secrecy rate is bounded by $C\left(\alpha_{\max}^2 P_1\right)$, which is bounded by a constant per the condition stated in Theorem 9. Hence the achieved secure degree of freedom remains the same as Theorem 4.